\documentclass{emulateapj}
\usepackage{epstopdf}
\usepackage{subfigure}

\begin{document}

\newcommand{\msun}{M_{\odot}}

\title{Gas Stripping in Simulated Galaxies with a multiphase ISM}
\author{Stephanie Tonnesen and Greg L. Bryan}
\affil{Department of Astronomy, Columbia University, Pupin Physics Laboratories, New York, NY 10027}

\begin{abstract}

Cluster galaxies moving through the intracluster medium (ICM) are expected to lose some of their interstellar medium (ISM) through ISM-ICM interactions.  We perform high resolution (40 pc) three-dimensional hydrodynamical simulations of a galaxy undergoing ram pressure stripping including radiative cooling in order to investigate stripping of a multiphase medium.  The clumpy, multiphase ISM is self-consistently produced by the inclusion of radiative cooling, and spans six orders of magnitude in gas density.  We find no large variations in the amount of gas lost whether or not cooling is involved, although the gas in the multiphase galaxy is stripped more quickly and to a smaller radius.  We also see significant differences in the morphology of the stripped disks.  This occurs because the multiphase medium naturally includes high density clouds set inside regions of lower density.  We find that the lower density gas is stripped quickly from any radius of the galaxy, and the higher density gas can then be ablated.  If high density clouds survive, through interaction with the ICM they lose enough angular momentum to drift towards the center of the galaxy where they are no longer stripped.  Finally, we find that low ram pressure values compress gas into high density clouds that could lead to enhanced star formation, while high ram pressure leads to a smaller amount of high-density gas.

\end{abstract}


\section{Introduction}

It has long been known that a galaxy's environment has an impact on its morphology:  spirals dominate in the field, and ellipticals and S0s dominate in dense cluster environments (Hubble {\&} Humason 1931).  This has been quantified in the density-morphology relation (Oemler 1974; Dressler 1980), but this relation alone does not determine whether the cluster environment affects the formation or the evolution of galaxies.  According to the Butcher-Oemler effect, clusters at z $\ge$ 0.2 are bluer and contain more spirals than nearby clusters, indicating that cluster environments may be correlated with the evolution of galaxies from spirals to earlier types (Butcher {\&} Oemler 1978).  More recently, by examining a specific merging cluster, Tran et al. (2005) presents a strong relationship between the Butcher-Oemler effect and infalling galaxies.

The environment-driven evolution of spiral galaxies into S0s is caused by mechanisms that physically remove gas and/or induce star formation until almost no gas remains.  Galaxy-intracluster medium (ICM) interactions are ones in which the gas in a galaxy interacts with the ambient intracluster hot gas.  One such process is ram pressure stripping (RPS), which removes the interstellar medium (ISM) of a galaxy as it moves through the ICM (Gunn {\&} Gott 1972).  Ram pressure could also compress the gas within a galaxy to cause a burst of star formation that would consume all of the remaining gas (Fujita {\&} Nagashima 1999).  A galaxy-ICM interaction does not affect the stellar component of a galaxy.  The cluster potential can also strip both gas and stars or compress gas to cause an increased star formation rate (Byrd {\&} Valtonen 1990).  These galaxy-cluster interactions affect both the stellar and gas components of galaxies.  There are also galaxy-galaxy interactions that can take place between galaxies.  These include mergers between galaxies with low relative velocities, and galaxy harassment: high-speed interactions between cluster galaxies (Hashimoto et al. 1998; Bekki 1999; Barnes {\&} Hernquist 1991; Bekki 1998; Moore et al. 1996).  These interactions can cause an increased star formation rate and will also affect both the stellar and gas component of galaxies.  

In our earlier work (Tonnesen, Bryan \& van Gorkom 2007), we studied cluster galaxies that had evolved in a cosmological simulation.  This simulation included dark matter and stellar particles, and followed gas through an adaptive mesh refinement (AMR) code.  It included radiative cooling and star formation.  This work was unique because it allowed the cluster to form in a cosmological simulation, but had the resolution to follow individual galaxies (3 kpc).  We found that most galaxies that lost all their gas did so without losing stellar mass, indicating that ram pressure stripping was the most effective evolutionary mechanism.  This convinced us to look more closely at ram pressure stripping of galaxies.  

While a substantial body of literature exists on ram pressure stripping (e.g. Roediger \& Hensler 2005; Roediger \& Br\"uggen 2006; Vollmer, Hutchmeier \& van Driel 2005; see references in van Gorkom 2004), the effect of a multiphase ISM has only begun to be explored in simulations.  Schulz and Struck (2001) included cooling and found that it caused a disk instability that led to an inflow of dense gas, and that including cooling allows the disk to remain cool and thin despite ICM heating.  Stripping of a multiphase ISM was also studied at higher resolution by Quilis, Moore {\&} Bower (2000).  They did not include radiative cooling, but instead randomly placed holes and overdensities in their ISM disk in an attempt to model an inhomogeneous ISM.  They find that stripping is much more complete as ICM streams through the holes in their galaxy and prevents stripped material from falling back onto the galaxy as earlier simulations had shown.  More recently, Kronberger et al (2008) have modeled cooling in the context of star formation.  They find increased star formation in the center of their galaxy and star formation in the stripped material (although it does not escape the galactic halo).

Observations have been inconclusive about the effect of stripping on dense molecular clouds.  In a survey of Virgo cluster galaxies, Kenney {\&} Young (1986, 1989) found that CO remains in HI deficient galaxies, indicating that molecular clouds are not being stripped entirely.  Whether this is because molecular clouds reside in the center of galaxies where even atomic gas is rarely stripped, or because molecular clouds are too dense to be stripped is unclear.   The likelihood of molecular gas being stripped varies with each individual case.  Clemens et al. (2000, 2001) have examined in detail the interacting galaxy pair NGC 4490 and NGC 4485.  They find evidence that in moving through the extended HI envelope surrounding the two galaxies, NGC 4485 has had atomic gas, molecular gas, and dust removed via ram pressure stripping.  Crowl et al. (2005) have studied the Virgo cluster galaxy NGC 4402, and observe ram pressure stripping and dense cloud ablation.  They find that it is possible to strip the lower density ISM without stripping dense clouds, but that the surviving clouds will eventually be ablated by the ICM wind.  They also observe an HII region which has likely formed within the stripped gas, so molecular gas was either stripped or formed from stripped material.  However, in NGC 4848, Vollmer et al (2001) find that the displaced molecular gas they observe is formed from re-accreted lower density gas, and that the molecular gas was not stripped at all.  NGC 4438 has been observed to have both stripped and surviving molecular gas, determined by the clouds' positions in the galaxy (Vollmer et al. 2005).  Another galaxy in the Virgo cluster, NGC 4522, has no HII regions in the outer disk, but extraplanar HII regions (Kenney {\&} Koopmann 1999; Kenney, van Gorkom {\&} Vollmer 2004).  

As discussed in Kenney et al (2004), there are four possibilities for stripping of molecular clouds.  (1) Molecular clouds can be directly stripped from the disk, although there is probably an upper limit to the surface density for which this is possible.  (2) Molecular clouds, which are thought to have a lifetime of about $10^7$ years (Blitz \& Shu 1980; Larson 2003, Hartmann 2003), can also evolve via star formation into lower density gas, which can then be stripped directly.   (3)  Another possibility is that the molecular clouds are not directly stripped out of the galaxy, but are instead ablated by the ICM wind on a longer timescale (Nulsen 1982).  (4)  Finally, dense clouds may be coupled to the rest of the ISM by magnetic fields or some other mechanism, so that the entire ISM is stripped out together.  

In this paper, we run a set of high resolution simulations (40 pc resolution, which is small enough to marginally resolve Giant Molecular Clouds) to understand how a multiphase ISM could affect ram pressure stripping of a galaxy's gas.  It is important to recognize that we do not attempt to include all of the physics involved in the ISM.  We use these simulations to understand how the density fluctuations that are observed in the multiphase ISM of galaxies will affect gas loss.  We vary ram pressure strength to investigate the result of different wind strengths.  Using our simulations, we are able to look in detail at possibilities (1) and (3).  We also consider whether including molecular clouds affects the large scale stripping properties such as the total amount of gas stripped and the radius to which gas is stripped.

In an introduction to our code, we provide the general characteristics of our simulation (\S 2.1-3).  We then discuss our suite of runs and the measurements we use to compare them (\S 2.4-5).  In \S 3 we discuss how cooling affects a galaxy evolving in a static ICM.  We then (\S 4) compare our simulations.  In \S 5 we briefly discuss the limitations and implications of our results.  Finally, we conclude in \S 6 with a summary of our results and the overall picture of ram pressure stripping of a galaxy with a multiphase ISM.  

\section{Methodology}

\subsection{Simulation}\label{sec-sim}

We have simulated a galaxy using the adaptive mesh refinement (AMR) code {\it Enzo}.  Although this code can follow stellar and dark matter particles, we chose to use a fixed stellar and dark matter potential for ease of computation, although the self-gravity of the gas is self-consistently computed.  To follow the gas, we employ an adaptive mesh for solving the fluid equations including gravity (Bryan 1999; Norman \& Bryan 1999; O'Shea et al. 2004).  The code begins with a fixed, static grid and automatically adds refined grids as required in order to resolve important features in the flow as defined by enhancements in the gas density.

The simulation includes radiative cooling using the Sarazin \& White (1987) cooling curve.  We allow cooling to 8,000 K, which results in gas with neutral hydrogen temperatures without overcooling a large fraction of our gas.  With this cutoff, we find that we still can form gas clouds with densities typical of molecular clouds, although it is clear that the internal structure of such clouds is not reproduced in detail (see Figure \ref{fig:mass_rho} for the density mass distribution in the gas disks of our galaxies).  We discuss this choice in greater detail in \S \ref{sec-cool}.

Our box is 311 kpc on a side.  Our coarsest resolution is 2.5 kpc.  We allow 6 levels of refinement for our runs that include radiative cooling, for a best resolution of 40 pc.  The runs without radiative cooling have 5 levels of refinement.  We refine our simulation using baryon mass, with our minimum overdensity for refinement set as 51.2, which we found immediately refined the entire galactic disk to our best resolution.  We discuss resolution effects in more detail in \S \ref{sec-res}.  Our galaxy is placed at (0.5,0.5,0.22) or (155.5,155.5,68.42) kpc, so that we can follow the stripped gas as far as possible.  The ICM wind flows along the z-axis in the positive direction, with the lower z boundary set for inflow and higher z boundary set as outflow.    The x and y boundaries are set to reflecting.

\subsection{The Galaxy}

We model a massive spiral galaxy with a flat rotation curve of 200 km $s^{-1}$.  It consists of a gas disk that is followed using adaptive mesh refinement (including self-gravity of the gas -- a crucial ingredient required to form self-gravitating molecular clouds), as well as the static potentials of a stellar disk, a stellar bulge, and a dark matter halo.  We directly follow Roediger \& Br\"uggen (2006) in our modeling of the stellar and dark matter potential and gas disk.  Briefly, we model the stellar disk using a Plummer-Kuzmin disk (see Miyamoto \& Nagai 1975), the stellar bulge using a spherical Hernquist bulge (Hernquist 1993), and the dark matter halo using the spherical model of Burkert (1995).  This dark matter halo model is compatible with observed rotation curves (Burkert 1995; Trachternach et al. 2008).  The equation for the analytic potential is in Mori \& Burkert (2000).  

The gas is described as a softened exponential disk:  
\begin{equation}
\rho(R,z) = \frac{M_{\rm gas}} {2\pi a^2_{\rm gas}b_{\rm gas}}0.5^2 
{\rm sech} \left( \frac{R}{a_{\rm gas}}\right)
{\rm sech} \left( \frac{|z|}{b_{\rm gas}}\right)
\end{equation}
Given this gas density distribution in the disk, the gas temperature and pressure are calculated to maintain the disk in hydrostatic equilibrium with the surrounding ICM in the z direction.  The gas disk's rotational velocity is set so that the combination of the centrifugal force and the pressure gradient of the disk balances the radial gravitational force.  We cut the gas disk smoothly by multiplying the gas density distribution by $0.5(1 + \rm cos(\pi (R - 20$ kpc$)/26(21)$ kpc$))$ for 20 kpc $<$ R $\leq$ 26(21) kpc.  See our galaxy parameters in Tables \ref{tbl-const} and \ref{tbl-gconst}.  We use different initial gas disk radii in order to more easily compare disks when the ICM wind hits the galaxy.  We discuss this in greater detail in \S \ref{sec-nowind}.

\begin{table}
\begin{center}
\caption{Galaxy Stellar and Dark Matter Constants\label{tbl-const}}
\begin{tabular}{c | c}
\tableline
Variable & Value\\
\tableline
M$_*$ & $1 \times 10^{11}$ M$_{\odot}$ \\
a$_*$ & 4 kpc \\
b$_*$ & 0.25 kpc\\
M$_{bulge}$ & $1 \times 10^{10}$ M$_{\odot}$ \\
r$_{bulge}$ & 0.4 kpc \\
r$_{DM}$ & 23 kpc \\
$\rho_{DM}$ & $3.8 \times 10^{-25}$ g cm$^{-3}$ \\
\tableline
\end{tabular}
\end{center}
\end{table}

\begin{table}
\begin{center}
\caption{Gas Disk Constants\label{tbl-gconst}}
\begin{tabular}{c | c | c}
\tableline
Variable & Value RC & Value NC\\
\tableline
M$_{gas}$ & $1 \times 10^{10}$ M$_{\odot}$ & $1 \times 10^{10}$ M$_{\odot}$ \\
a$_{gas}$ & 7 kpc & 6.5 kpc \\
b$_{gas}$ & 0.4 kpc & 0.4 kpc\\
\tableline
\end{tabular}
\end{center}
\end{table}

\subsection{ICM Conditions}\label{sec:ICM}

Our galaxies evolve in a pressurized ICM.  Because a galaxy moving through the ICM can be more easily simulated by modeling a fixed galaxy within a moving ICM, our galaxy remains in the same place in our simulated box (at least the stars and dark matter do).  The galaxy initially evolves in a static ICM, to examine the stability in the static ICM and to allow cool, dense gas to form.  Later, we trigger a constant ICM inflow along the z-axis, which is always face-on to the galaxy.  See the Appendix for exactly when the wind hits each galaxy.  Briefly, the wind is triggered so that it hits the galaxy while gas is still distributed out to a large radius, but enough time has passed so that high density gas clouds have formed.  The exact time was chosen to be when the galaxy in the corresponding RCNW case had gas collapsed to $\rho$ $\geq$ 10$^{\rm -20}$ g cm$^{-3}$, typical of densities found in molecular clouds.

\begin{figure}
\includegraphics[scale=0.48]{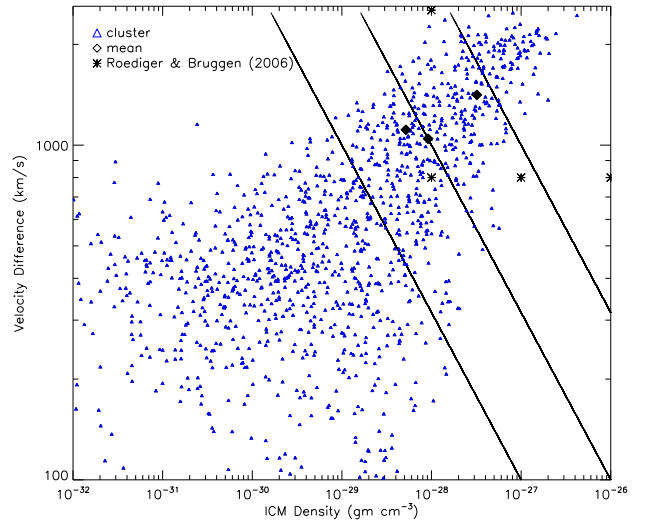}
\caption{The measurements of the ICM density and difference in velocity between the galaxies and the ICM for:  the cluster in the large simulation we examined in an earlier paper (Tonnesen, Bryan \& van Gorkom 2007, as blue triangles), the mean values we used for our three detailed simulations (black diamonds), and the values used by Roediger and Br\"uggen (2006) (black asterisks).  To guide the eye, we also plot lines of constant ram pressure at $1 \times 10^{-11}$ dynes cm$^{-2}$, $1 \times 10^{-12}$ dynes cm$^{-2}$, and $1 \times 10^{-13}$ dynes cm$^{-2}$.  See \S \ref{sec:ICM} for discussion.}
\label{fig:sample_select}
\end{figure}

We choose to study three ram pressure strengths using realistic values for the ICM density and velocity. The three ram pressure strengths we choose are $P_{\rm ram} = \rho v^2_{\rm ICM} = 6.4 \times 10^{-12}$ dynes cm$^{-2}$, $1 \times 10^{-12}$ dynes cm$^{-2}$, and $6.4 \times 10^{-13}$ dynes cm$^{-2}$ (see Table~\ref{tbl-ICM}).  We choose the maximum and minimum ram pressure strengths to match two of the values used by Roediger and Br\"uggen (2006), and we pick $1 \times 10^{-12}$ dynes cm$^{-2}$ because it is the highest ram pressure value at the virial radius of our cluster simulation (Tonnesen et al. 2007).

To choose the corresponding ICM parameters, $\rho$ and $v_{\rm ICM}$, we use the results from our earlier cluster simulation to find the mean density, velocity, and temperature of the ICM at the three ram pressures (Tonnesen, Bryan \& van Gorkom 2007).  This is shown in Figure \ref{fig:sample_select}, where for comparison we plot the values used by Roediger \& Br\"uggen (2006).  As discussed earlier, we set the inflow boundary condition in order to model the ICM wind.  However, because we need to allow our galaxies to cool and evolve before the ICM wind begins to strip them, the galaxies are initialized in a static ICM and then, after a multiphase ISM has developed, the boundary values are set to generate a wind with the desired characteristics.  To get the parameters for this initial ICM we backtrack from the density, temperature and velocity of the ICM wind using the Rankine-Hugoniot jump conditions, assuming a Mach number of 3.5.  We choose this Mach number because it gives us a supersonic shock (which leads to cleaner behavior at the inflow boundary), while allowing our initial static ICM to have a similar density and temperature as the ICM wind.  This procedure is important because we compare the evolution of our stripped galaxy to one that only cools in a static medium that has the same density and temperature as the flowing ICM.  We use the same density and temperature as in the moving ICM for the cases in which the ICM always remains static, because the stripped galaxies spend more time in the simulation in the moving ICM.

\begin{table*}
\begin{center}
\caption{ICM Data\label{tbl-ICM}}
\begin{tabular}{c | ccccccc}
\tableline
Runs & $\rho_{ICM}$ & T$_{ICM}$ & P$_{thermal}$ & P$_{ram}$ & v$_{ICM}$ & Cooling? & r$_{initial}$\\
\tableline
PHRCW & $3.20 \times 10^{-28}$& $ 4.01 \times 10^7$ & $1.765 \times 10^{-12}$ & $6.4 \times 10^{-12}$ & 1413.5 & Y &25 \\
PHNCW &  $3.20 \times 10^{-28}$ & $ 4.01 \times 10^7$ & $1.765 \times 10^{-12}$ & $6.4 \times 10^{-12}$ & 1413.5 & N &21\\
PHRCNW &$3.20 \times 10^{-28}$ & $ 4.01 \times 10^7$ & $1.765 \times 10^{-12}$ & 0.0 & 0.0 & Y &25\\
\tableline
PMRCW & $9.15 \times 10^{-29}$& $2.52 \times 10^7$ & $3.169 \times 10^{-13}$ & $1 \times 10^{-12}$ & 1045.4 & Y &25\\
PMNCW & $9.15 \times 10^{-29}$ & $2.52 \times 10^7$ & $3.169 \times 10^{-13}$ & $1 \times 10^{-12}$ & 1045.4 & N &21\\
PMRCNW & $9.15 \times 10^{-29}$& $2.52 \times 10^7$ & $3.169 \times 10^{-13}$ & 0.0 & 0.0 & Y &25\\
\tableline
PLRCW & $5.18 \times 10^{-29}$& $2.19 \times 10^7$ & $1.557 \times 10^{-13}$ & $6.4 \times 10^{-13}$ & 1111.7 & Y &25\\
PLNCW & $5.18 \times 10^{-29}$ & $2.19 \times 10^7$ & $1.557 \times 10^{-13}$ & $6.4 \times 10^{-13}$ & 1111.7 & N &21\\
PLRCNW & $5.18 \times 10^{-29}$& $2.19 \times 10^7$ & $1.557 \times 10^{-13}$ & 0.0 & 0.0 & Y &25\\
\tableline
\tableline
\end{tabular}
\end{center}
\end{table*}

\subsection{Suite of Simulations}

We perform nine full runs because while attempting to understand how cooling affects galactic gas stripping, we also need to understand how cooling alone affects a galaxy.  We examine how three different ram pressure strengths affect galaxies that include and do not include radiative cooling.  We choose a naming convention for the simulations which indicates the parameters used for that simulation (see also Table~\ref{tbl-ICM}).  The three ram pressure strengths also correspond to the three ICM thermal pressures, and are denoted PH, PM, PL for high, medium, and low.  If cooling is turned on, the next identifier is RC, if not, NC.  The six runs with an ICM wind end with W.  In addition to these six runs, we also study three runs of galaxies evolving with radiative cooling in a static ICM with the same temperature and density as the moving ICM.  All these runs end with NW, for no wind.  For details on these nine simulations, see Table \ref{tbl-ICM}, as well as the appendix.  Because we are interested primarily in the impact of radiative cooling on gas stripping, we only consider cases with a face-on wind.  These cases are likely to be the most affected by an ICM wind because the gas density fluctuations in the disk, caused by radiative cooling, are most apparent in a face-on view.

\subsection{The Measurements}

We look closely at three types of measurements: 1) the total amount of gas in the galaxy, 2) the radius of the dense gas in the galaxy, and 3) the amount of gas mass at different densities. 

We first measure the total amount of gas that remains in the galaxy.  We measure this in two ways:  the amount of gas mass in a cylinder centered on the galaxy with a height of 10 kpc and a radius of 27 kpc, and the total amount of bound gas mass in the entire box.  Mass is bound if the thermal and kinetic energy are smaller than the gravitational energy from the static potential of the stars and dark matter.  We can compare these values to an analytic estimate of how much gas a galaxy should lose using two analytic mass loss estimates, corresponding to two stages of gas stripping.  We will follow Roediger \& Br\"uggen (2007) by naming the initial stage ram pressure pushing and the later stage continuous stripping.  We can predict the amount of gas loss from ram pressure pushing by measuring the amount of gas mass outside the analytic Gunn \& Gott (1972) stripping radius, the radius at which the restoring force per unit area from the galaxy is equal to the ram pressure acting on the galaxy, or where $\rho  v^2 =  2 \pi G \Sigma_{\rm star} \Sigma_{\rm gas}$.  We calculate the radius using the initial gas distribution and measure the mass outside this radius immediately before the ICM wind hits the galaxy.  Once initial gas stripping has occurred, there is a slower continuous stripping phase which is caused by the Kelvin-Helmholtz instability.  Nulsen (1982) derived an equation for the mass loss rate of a spherical cloud, which Roediger \& Br\"uggen (2007) adapted for a disk:  $\dot{M}_{\rm KH} = 0.5\pi R^2 \rho_{\rm ICM}v_{\rm gal}$.  Because we use a constant density and velocity ICM wind, only the changing radius of the galaxy will change this rate of gas loss.

We measure the stripping radius in the simulation by finding the radius of gas with a density $\rho > 10^{-26}$ g cm$^{-3}$ (or approximately a number density of $n \sim 10^{-2}$ cm$^{-3}$).  To mitigate the effect of asymmetry on this measurement, we find the radius in 12 azimuthal slices around the disk and take the average.  Because cooling causes both overdensities and underdensities, we measure this radius by finding the distance to the furthest cell with gas above our threshold density.  As may be expected, this increases the radius we measure for the galaxies that include cooling, and has no effect on the radius of the galaxies without cooling.  We compare this to the analytic stripping radius.

We also measure the amount of gas at different densities.  The gas is split into three density regimes:  (1) high density: $\rho > 10^{-20}$ g cm$^{-3}$, (2) moderate density: $10^{-20}$ g cm$^{-3} > \rho > 10^{-22}$ g cm$^{-3}$, and (3) low density: $ 10^{-22}$ g cm$^{-3} > \rho > 10^{-26}$ g cm$^{-3}$.  We choose these density ranges in order to follow three different types of gas: gas with high densities seen in molecular clouds (Crutcher 1991 and references therein), gas with densities that are nearly that of molecular gas, and finally gas which is clearly not molecular, but more dense than the surrounding ICM.  We use the gas that is within the cylinder centered on our galaxy for this measurement.

\section{Galaxies without an ICM wind}\label{sec-nowind}

\begin{figure}
\includegraphics[scale=0.5]{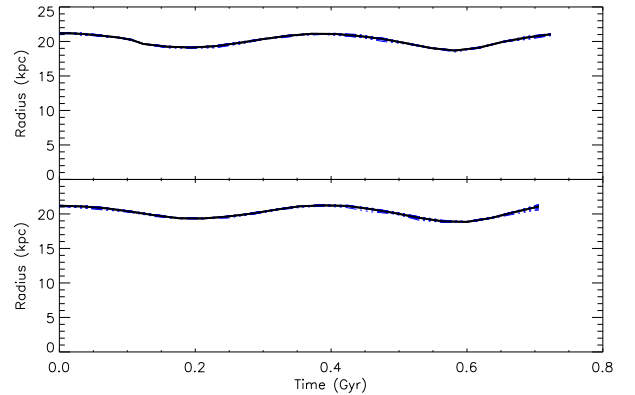}
\caption{The radii of two galaxies with no cooling and in a static ICM at two different pressures.  The radii fluctuate slightly at the epicyclic frequency.  We believe this is because of the precipitous pressure decrease at the edge of the galaxy.}
\label{fig:ncnw}
\end{figure}

Although the stability of this galaxy model is discussed in detail in Roediger \& Hensler (2005) and Roediger \& Br\"uggen (2006), it is important to reiterate that using {\it Enzo}, the galaxies are stable if there is no radiative cooling and no ICM wind.  Although they maintain the exact same amount of gas throughout the simulation, there is a fluctuation in the radius with the epicyclic frequency for this galaxy.  This is most likely due to the rapid change in gas density, and therefore pressure, in the outer kpc (between 20-21 kpc) of the galaxy in the initial galactic profile.  In Figure \ref{fig:ncnw}, we show how the radius varies with time for two of the runs in which the galaxy sits in a static environment with ICM pressure and without radiative cooling.

In the simulations in which we include radiative cooling, at all ICM pressures, gas with a density of $10^{-20}$ g cm$^{-3}$ forms.  At this point, gas with the density of molecular clouds has formed, and we consider the gas disk to be a multiphase ISM.  In all cases, the highest density gas ($\rho \ge$ 10$^{-20}$ g cm$^{-3}$) only forms stably within the central 7 kpc of the disk.  Also, the mass fraction of gas with densities of molecular gas ($\rho \ge$ 10$^{-21}$ g cm$^{-3}$) is 38\% - 50\% after evolving for $\ge$ 1 Gyr in a static ICM with radiative cooling.  We closely examine three galaxies with radiative cooling in three different pressure ICM's and no wind (the RCNW cases in Table \ref{tbl-ICM}).  

We find that -- as in the galaxies without cooling -- there does seem to be a radial epicyclic oscillation.  However, in the higher ICM pressure cases (PHRCNW and PMRCNW), low density and low pressure regions in the disk allow non-rotating ICM gas to enter and interact with the rotating denser clouds.  These clouds therefore slow and lose angular momentum, leading the disk to settle towards the center.  This drop in disk radius can be seen in the bottom panels of Figure \ref{fig:radii}.  It is notable that in the lowest pressure ICM there is very little entrainment of the ICM gas, and the radius of the gas is dominated by epicyclic oscillations.  We also find evidence of this process, and how it varies with ICM pressure, by following the total amount of gas in these galaxies.  For example, PHRCNW gains $3.8 \times 10^8$ M$_\odot$ over the 950 Myr run, while PLRCNW gains $2.1 \times 10^8$ M$_\odot$, because of the different amounts of gas entrained through cooling of the disk.  

Because the radius of the gas disk decreases as the gas cools (see the bottom panels of Figure \ref{fig:radii}), we lower the gas radius of the galaxies {\it without} cooling to the radius that the radiatively cooled galaxies reach due to their entrainment of the ICM (as described in the previous paragraph).  Therefore, although the radius of the galaxies with cooling begins at 25 kpc, the radius of the galaxies without cooling start at 21 kpc.  Despite this different initial radius, we set the radial gas scale length, a$_{gas}$, so that the galaxies with and without cooling have the same amount of mass.  This is all done in order to ensure a fair comparison between the radiatively cooling and non-cooling runs.

\section{The Effects of Ram Pressure on a Multiphase ISM}

\subsection{Morphology of the Stripped Disk}

\begin{figure}
\subfigure{\includegraphics[scale=0.1]{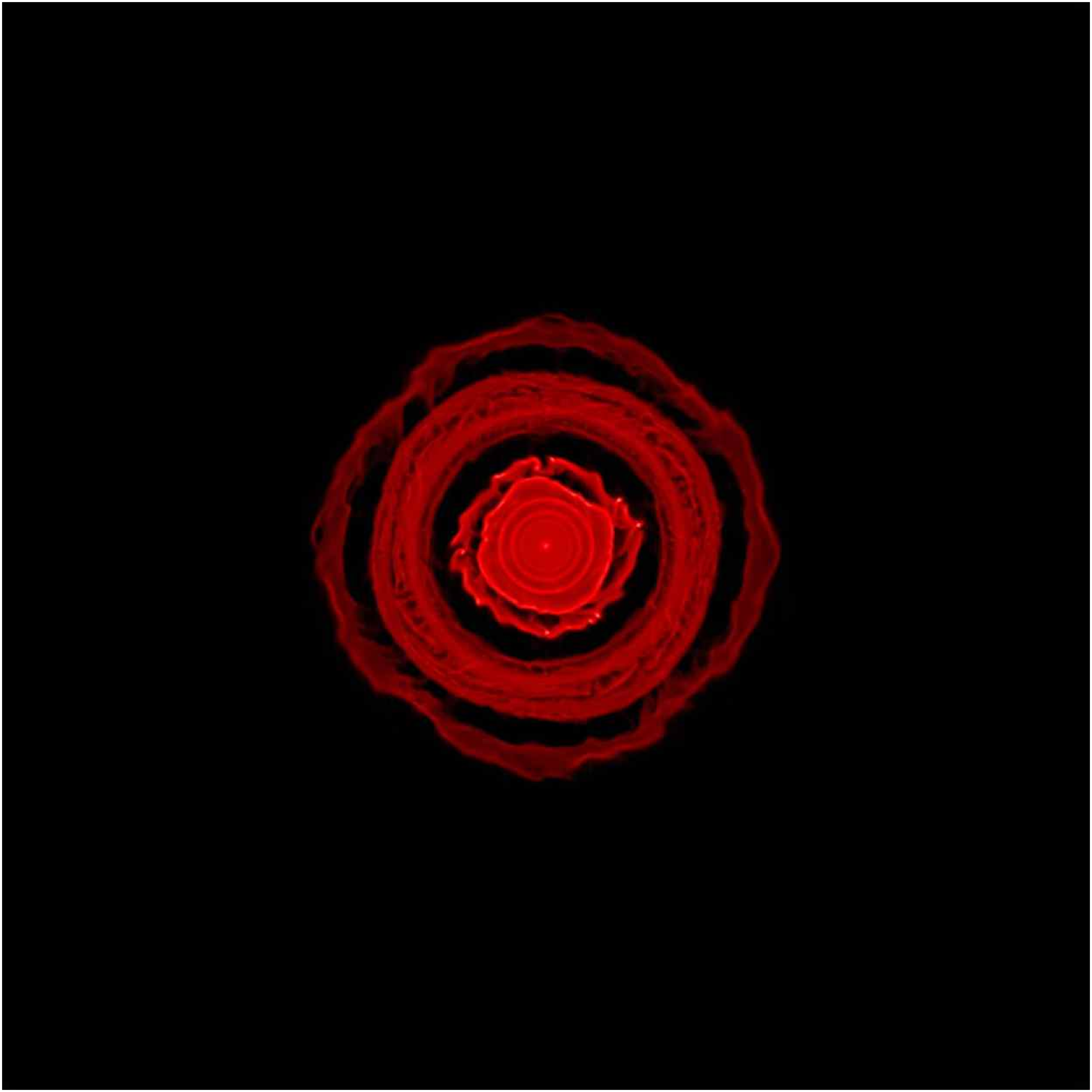}}
\subfigure{\includegraphics[scale=0.1]{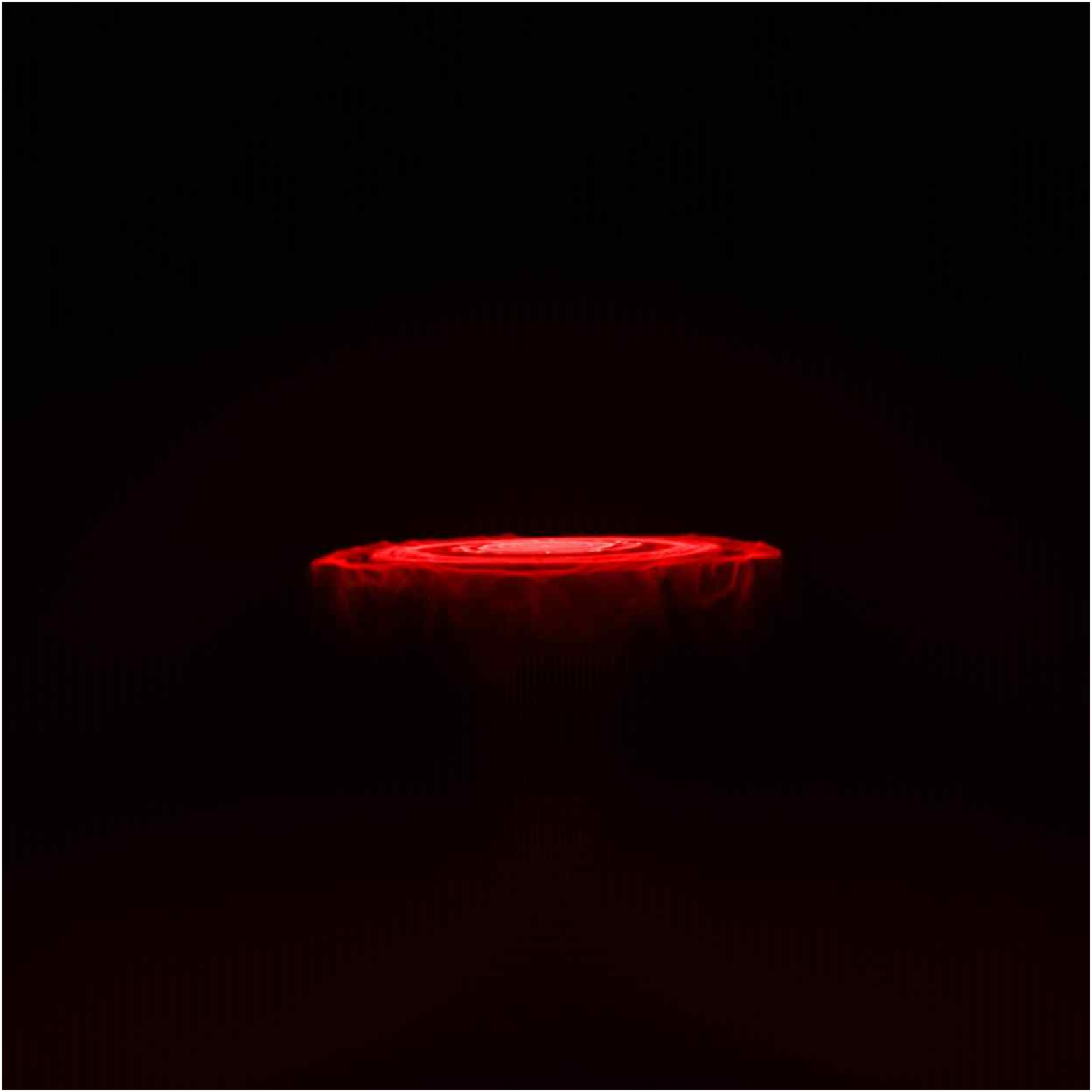}}\\
\subfigure{\includegraphics[scale=0.1]{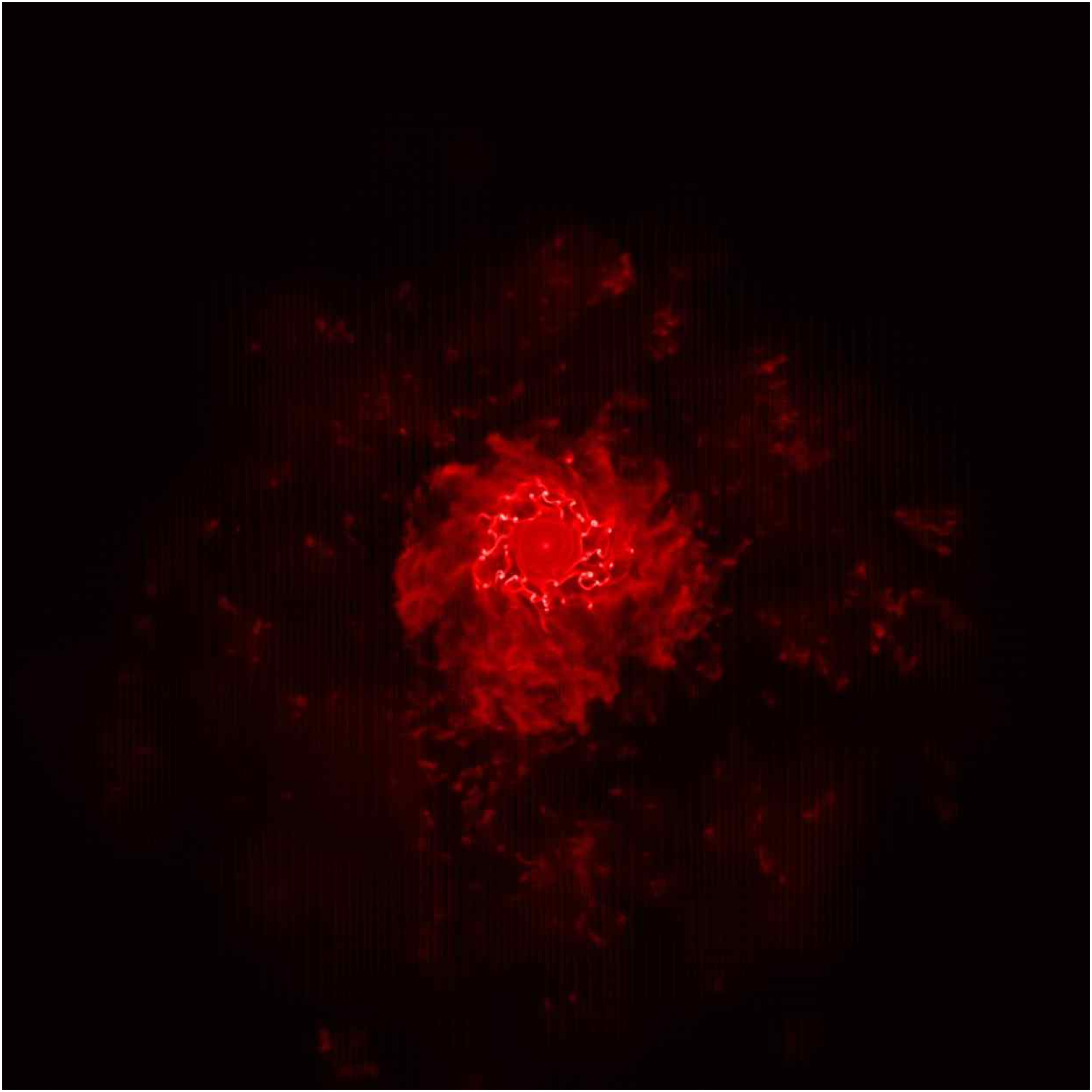}}
\subfigure{\includegraphics[scale=0.1]{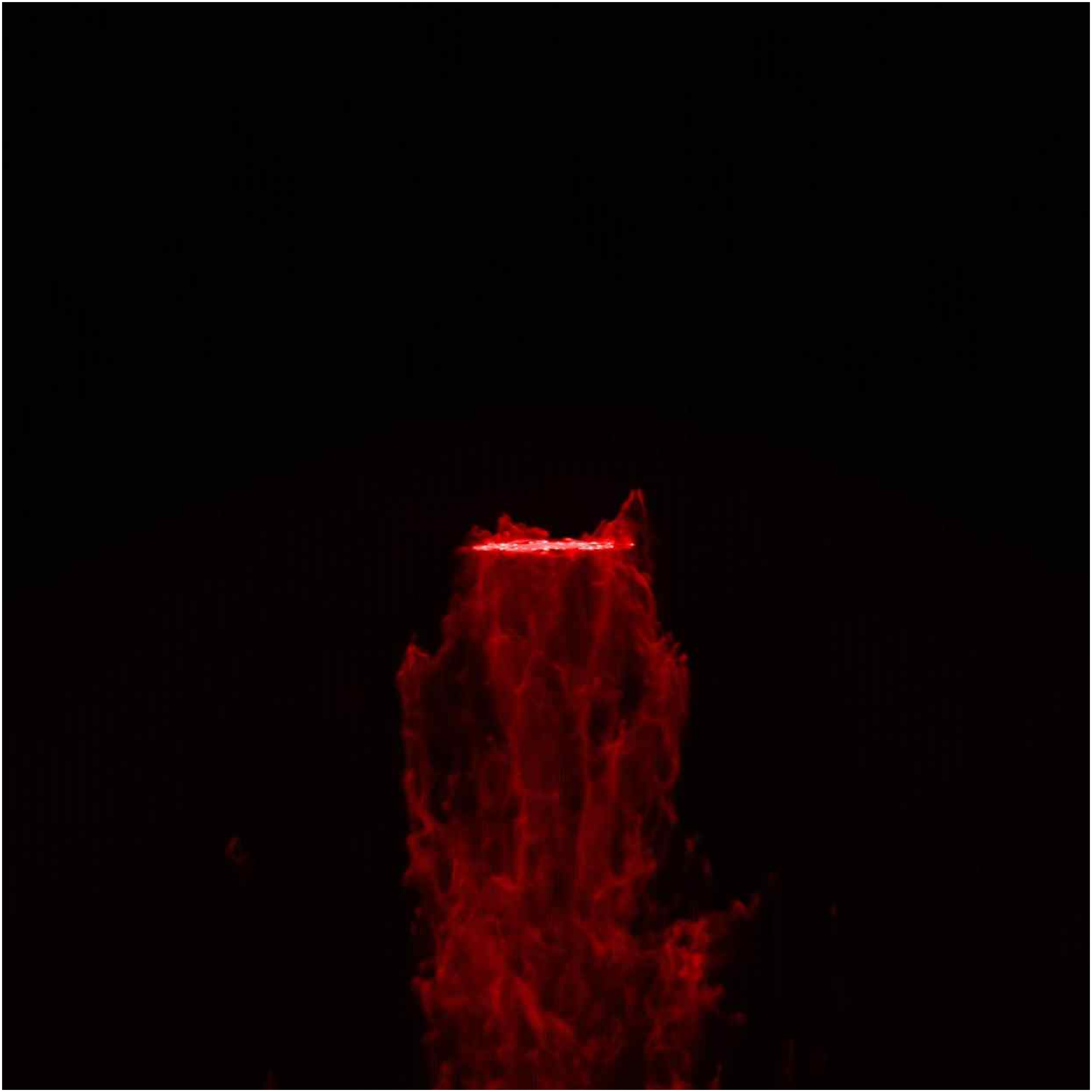}}\\
\subfigure{\includegraphics[scale=0.1]{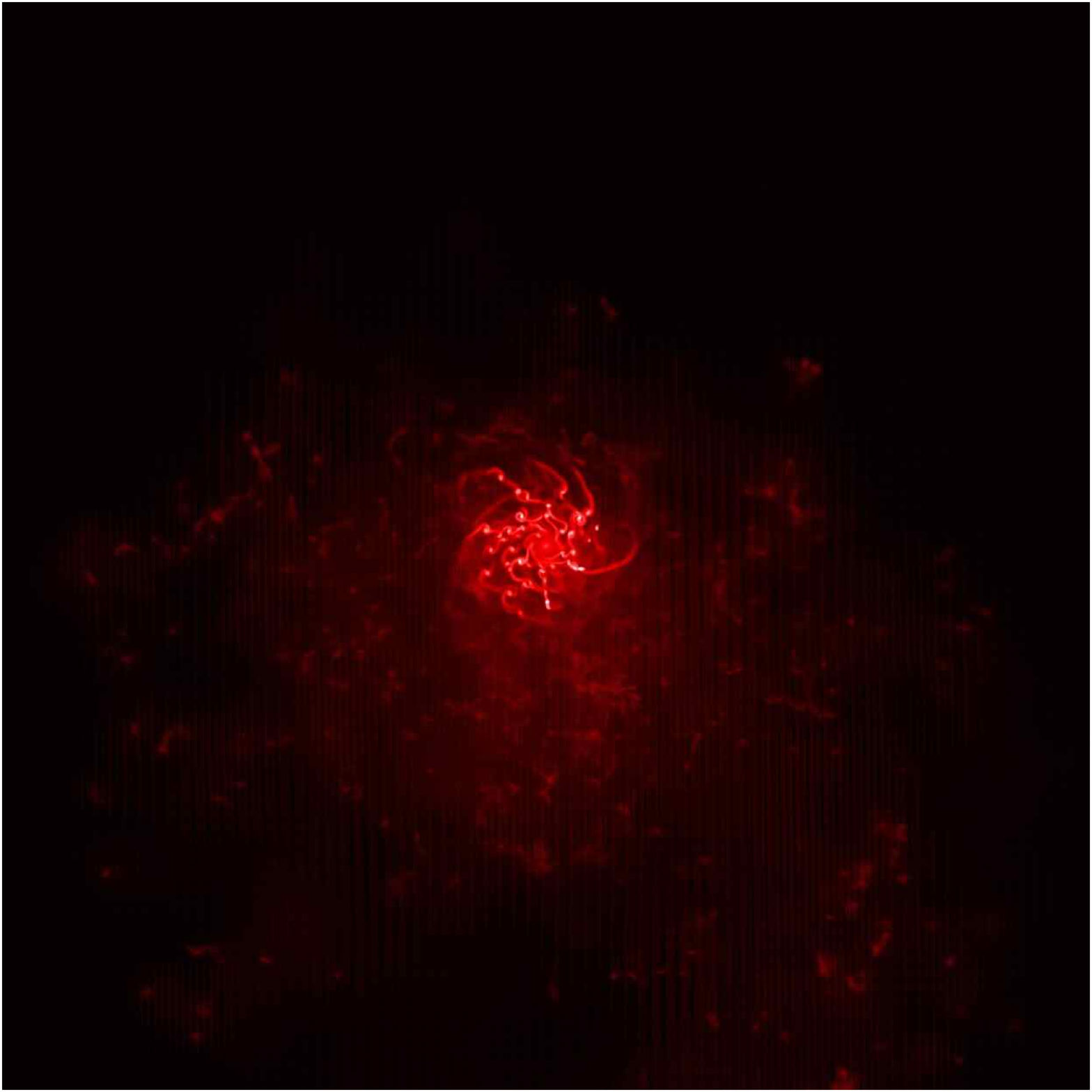}}
\subfigure{\includegraphics[scale=0.1]{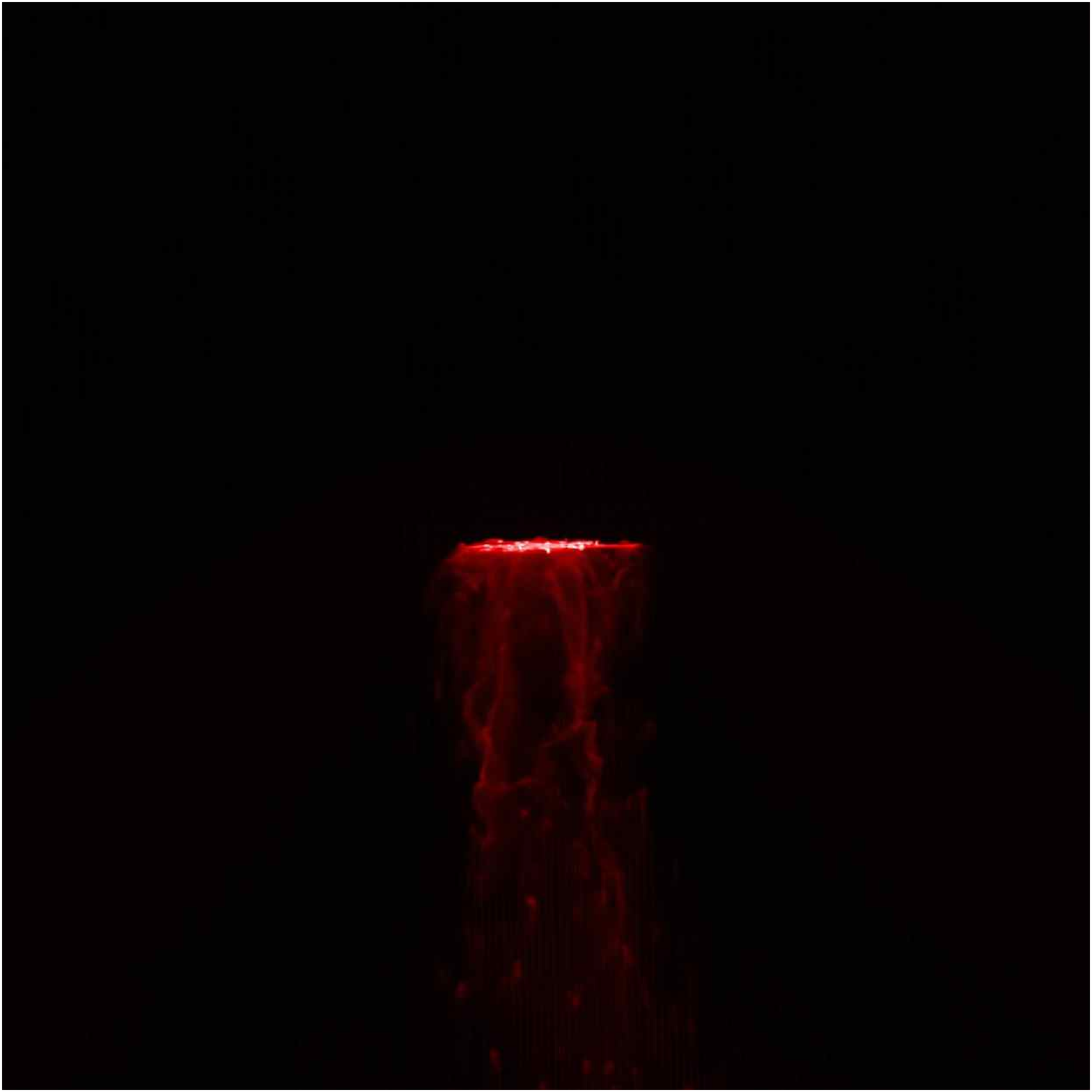}}\\
\subfigure{\includegraphics[scale=0.1]{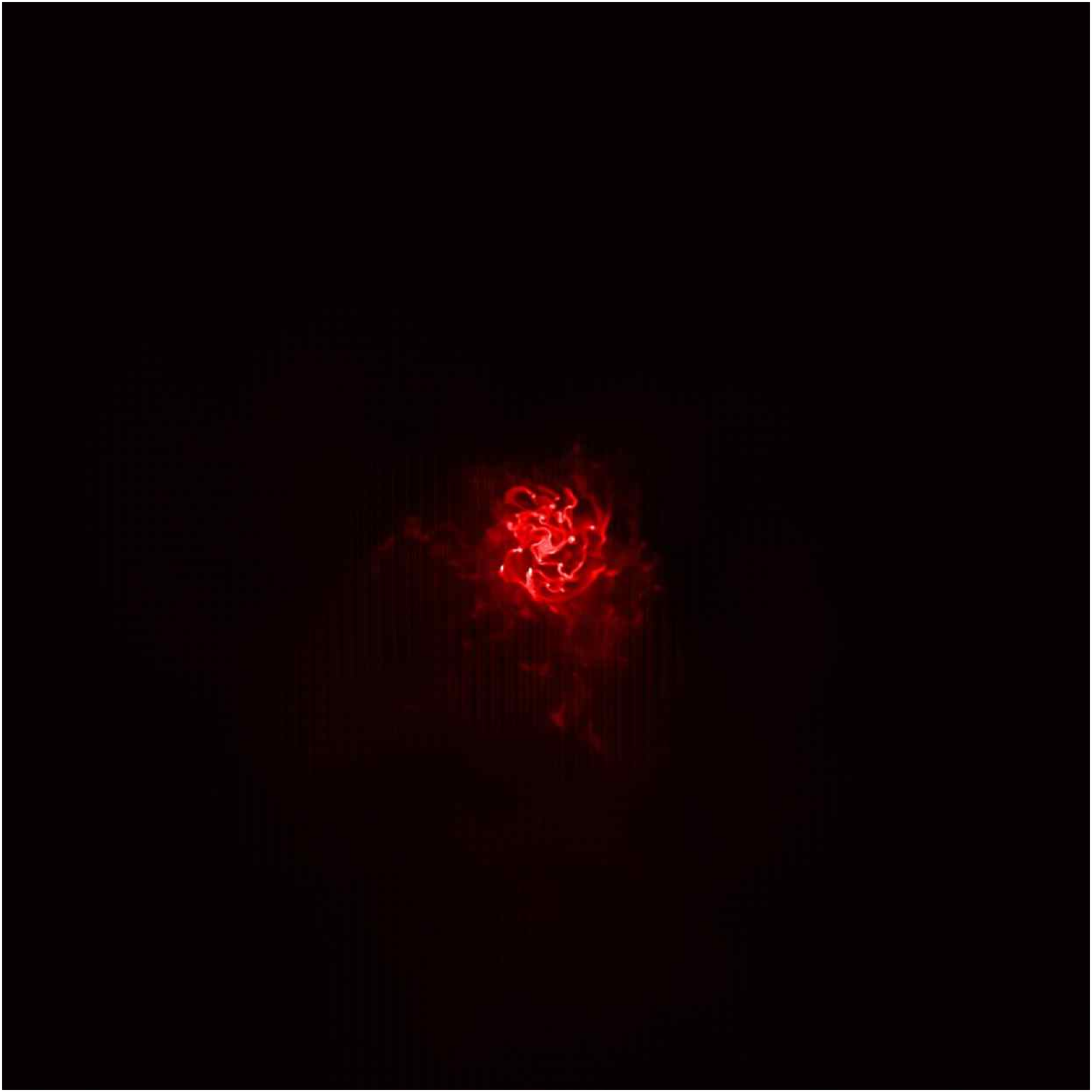}}
\subfigure{\includegraphics[scale=0.1]{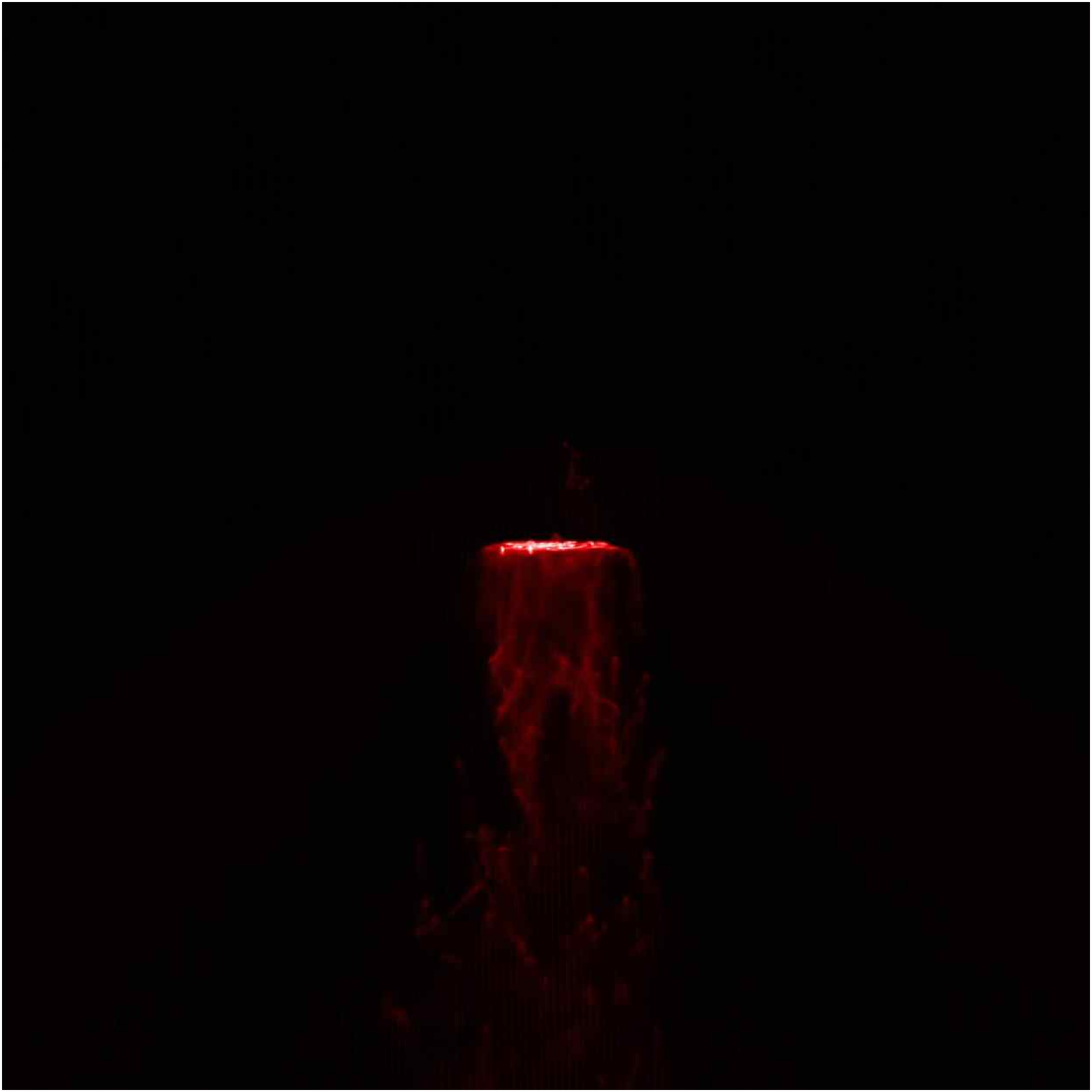}}\\
\caption{A face-on and edge-on view of gas surface density in simulation PHRCW, at 40, 250, 500, and 750 Myr after the wind has hit the galaxy.  Note the substructure both in the disk and in the stripped gas.  Each image side is 103.7 kpc, and the color scheme has a logarithmic stretch. }
\label{fig:rcshots}
\end{figure}

\begin{figure}
\subfigure{\includegraphics[scale=0.1]{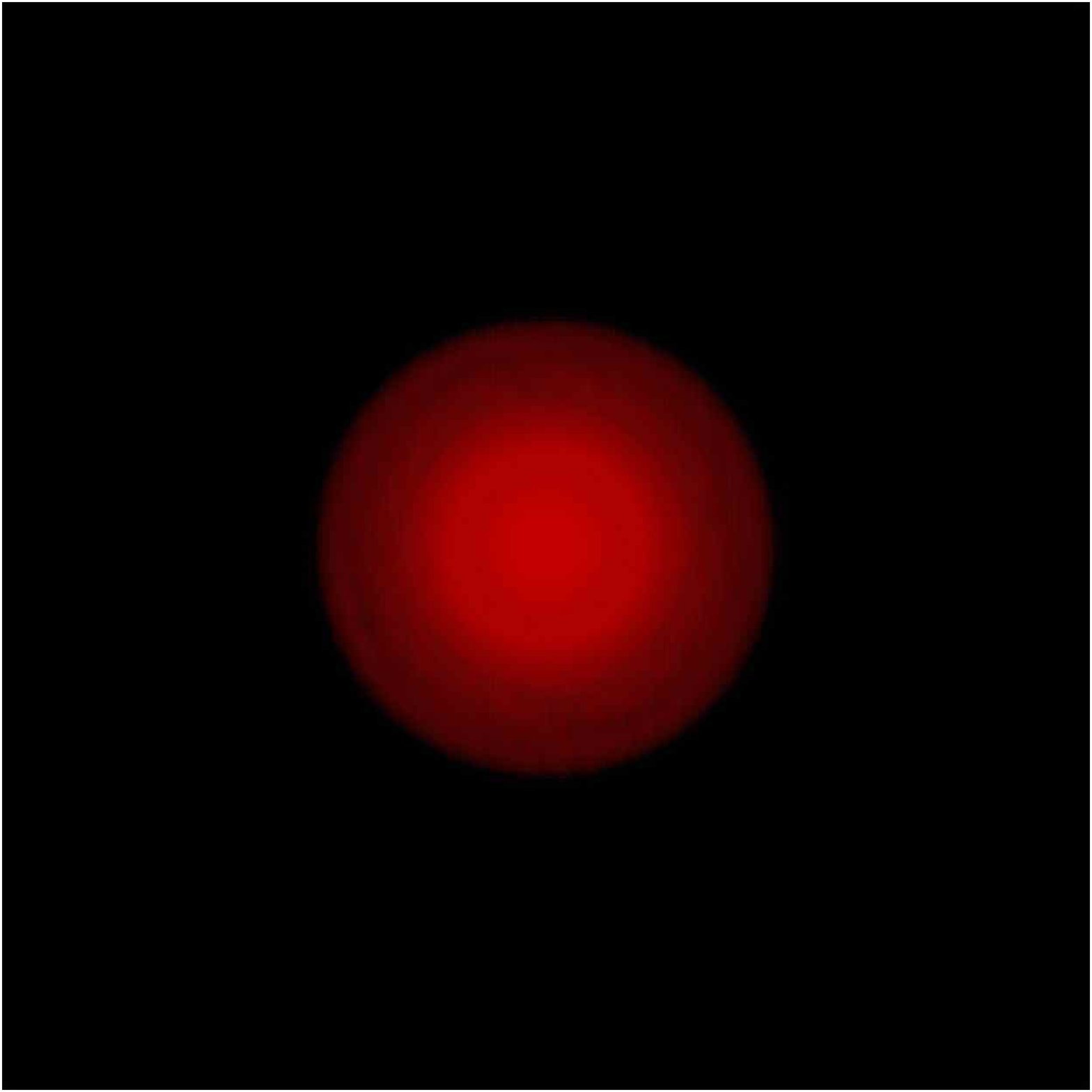}}
\subfigure{\includegraphics[scale=0.1]{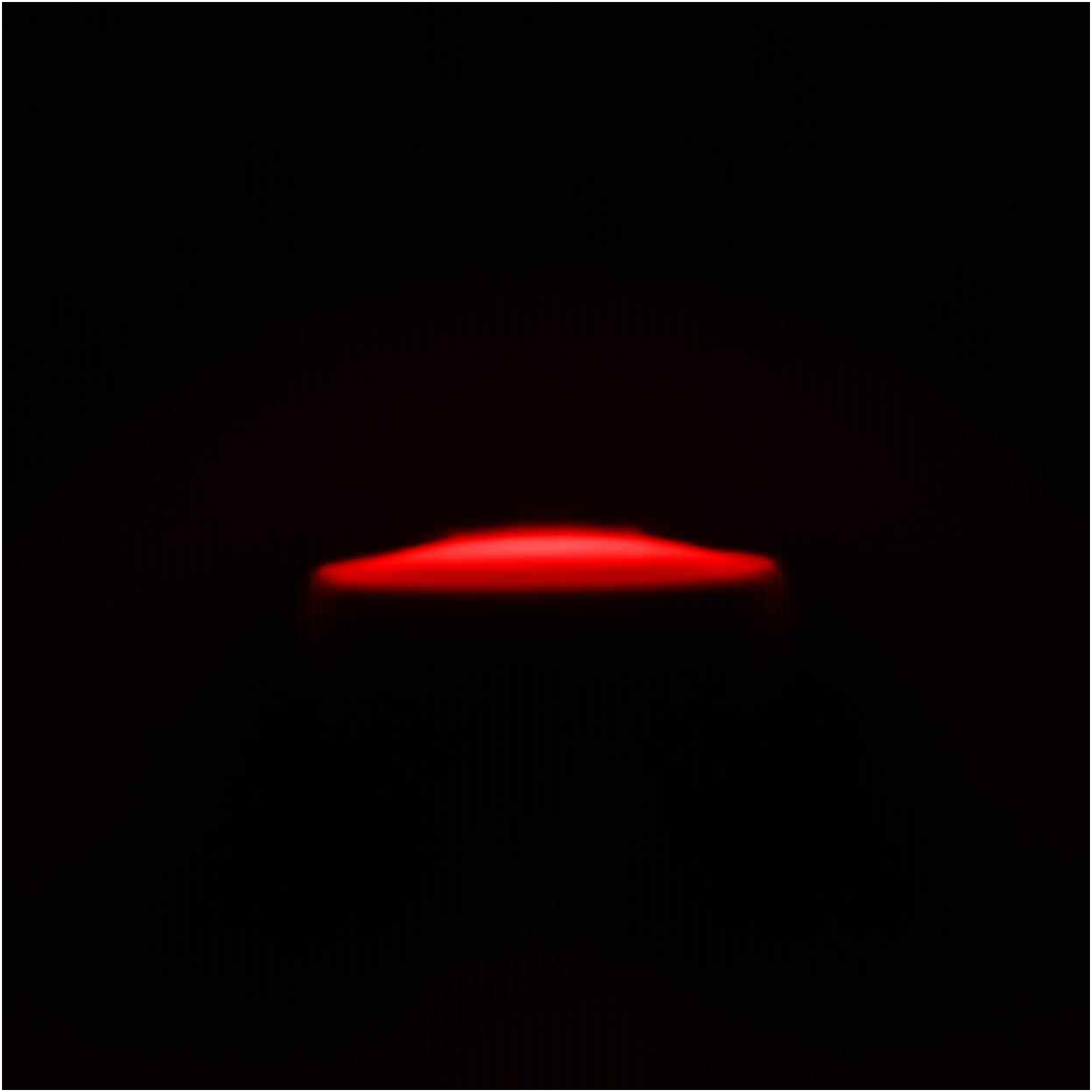}}\\
\subfigure{\includegraphics[scale=0.1]{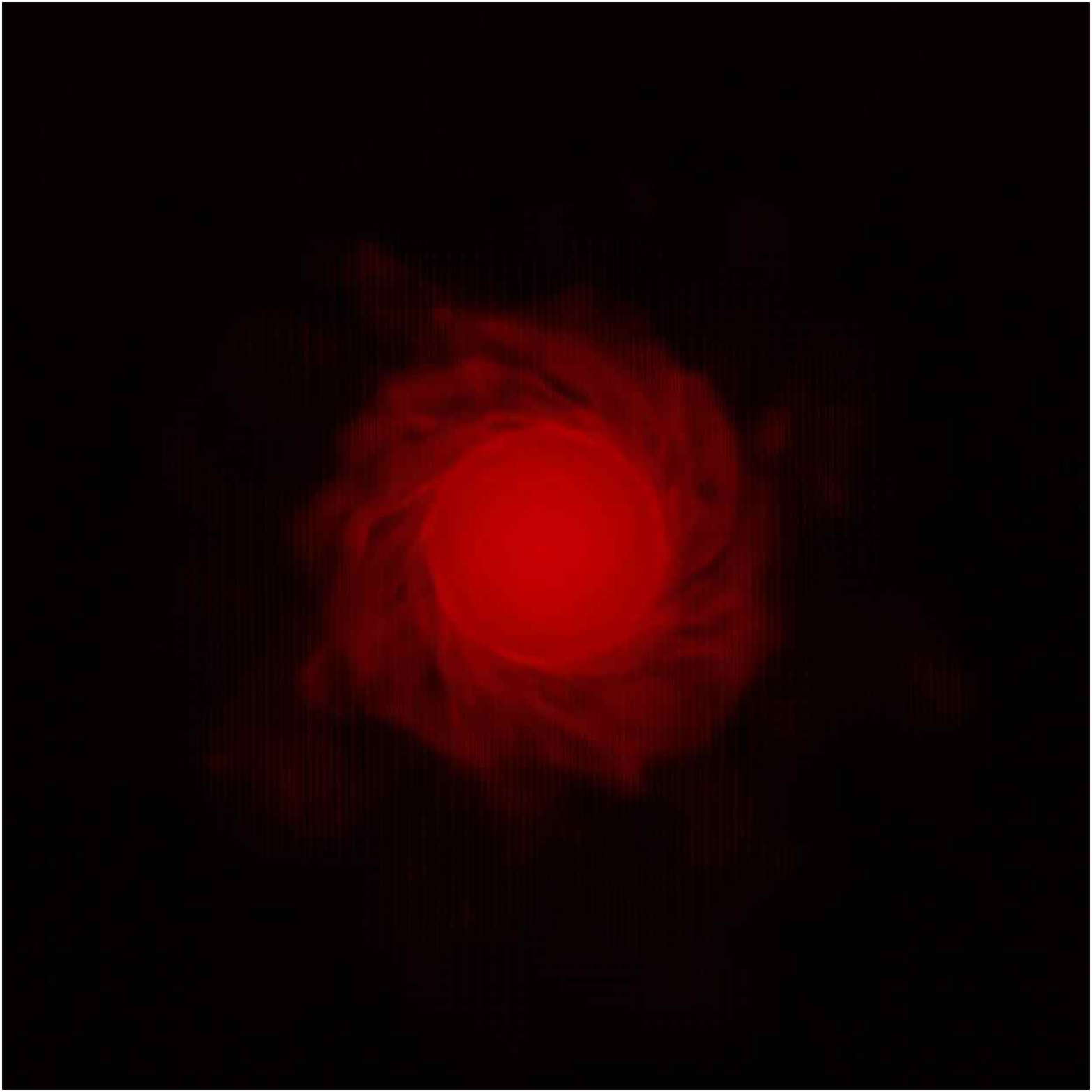}}
\subfigure{\includegraphics[scale=0.1]{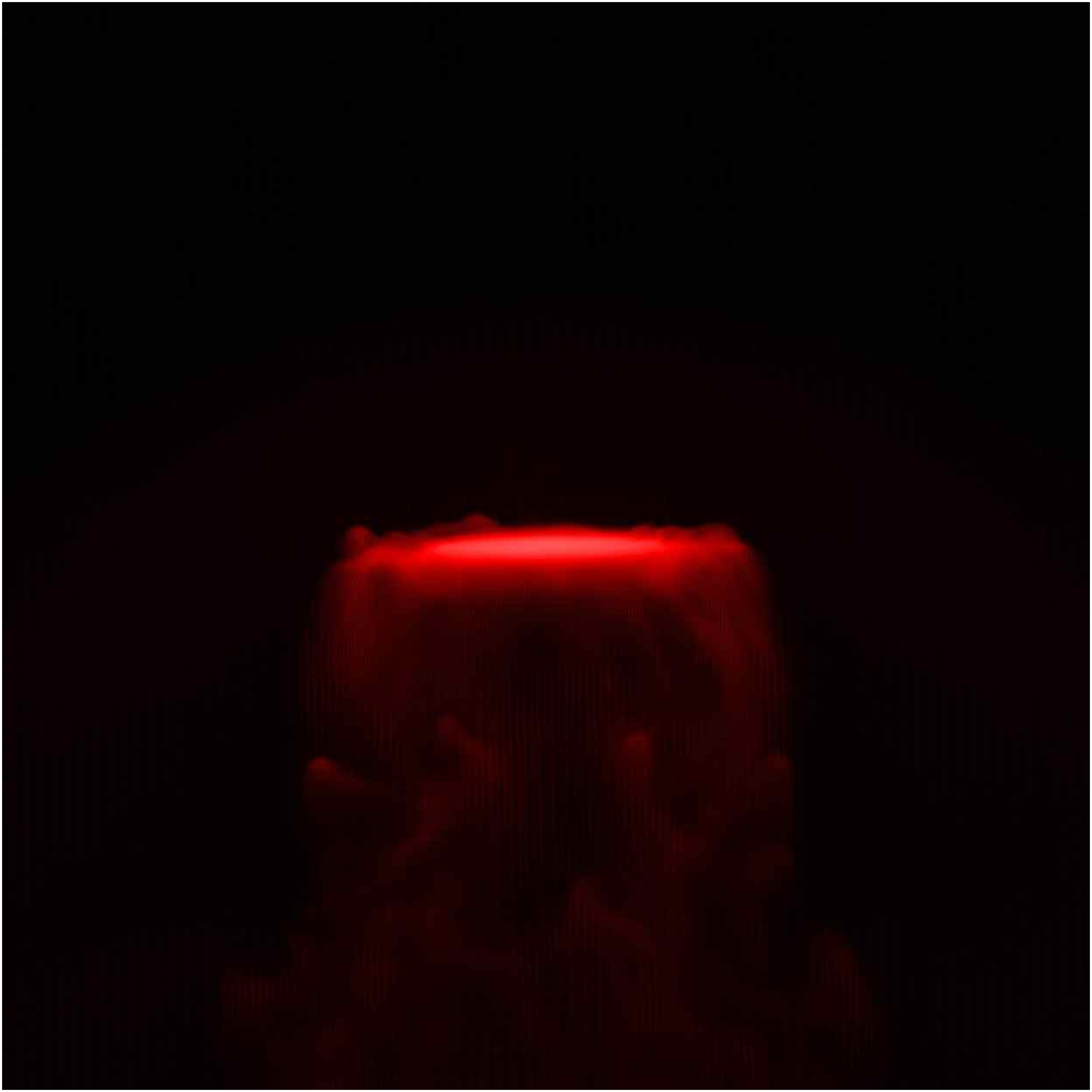}}\\
\subfigure{\includegraphics[scale=0.1]{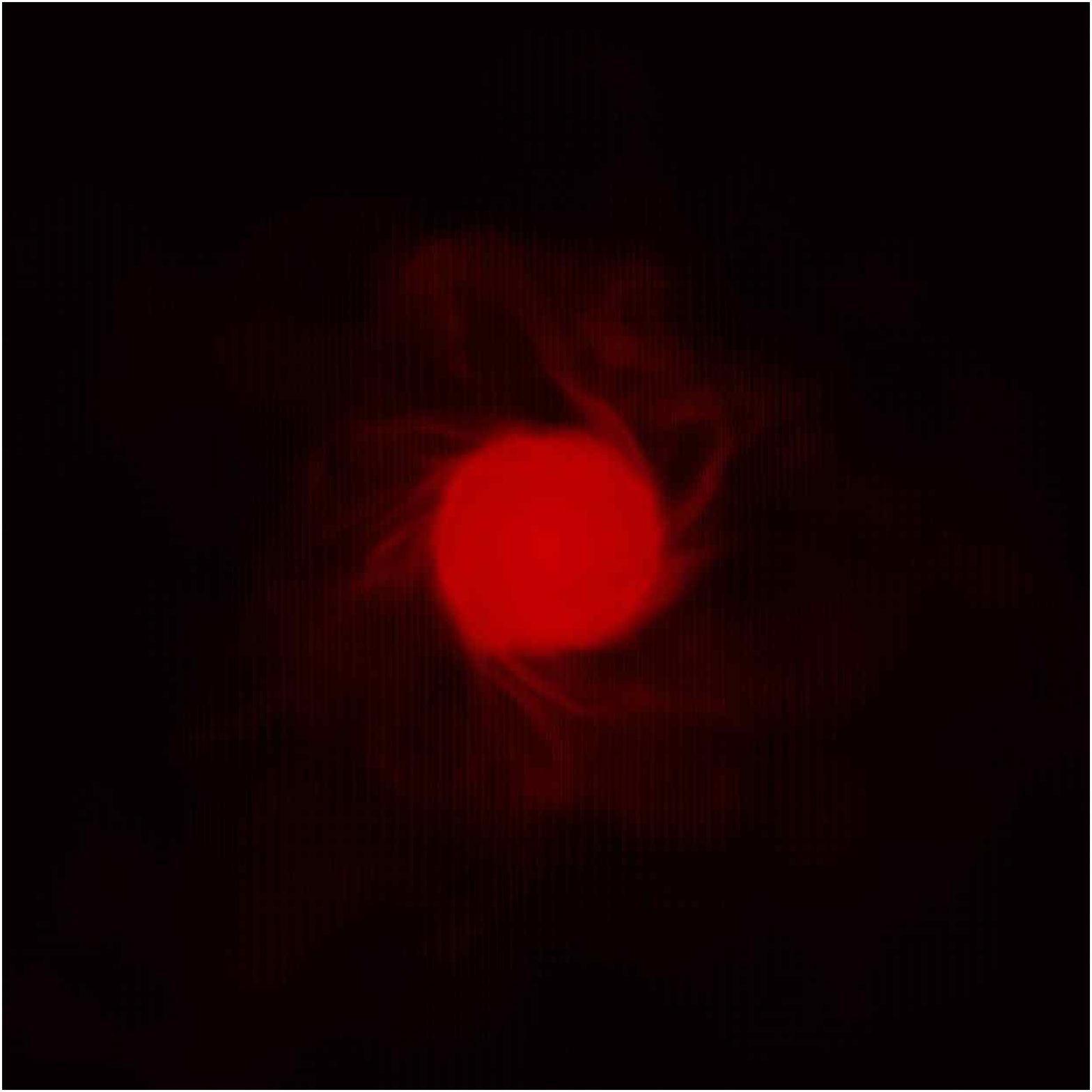}}
\subfigure{\includegraphics[scale=0.1]{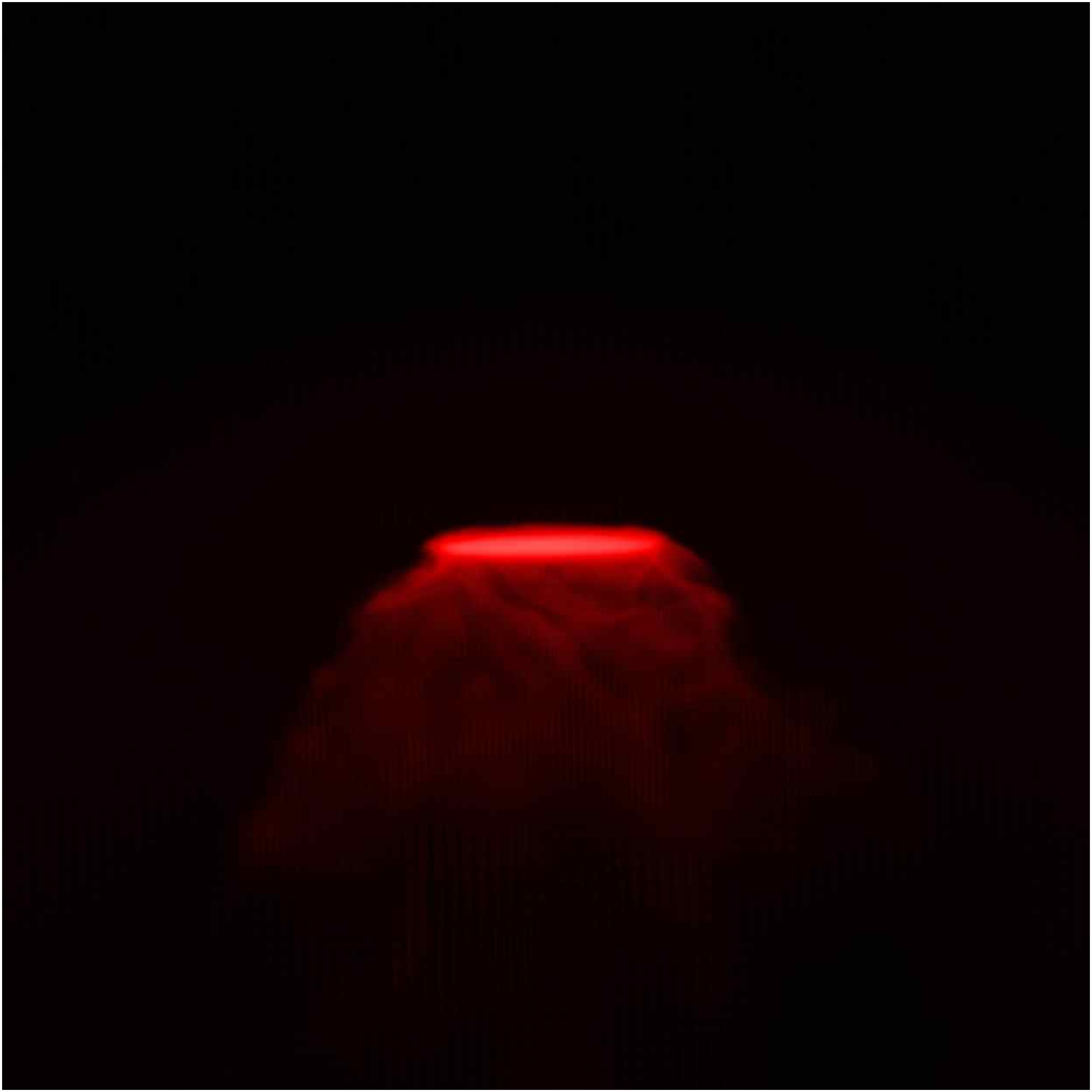}}\\
\subfigure{\includegraphics[scale=0.1]{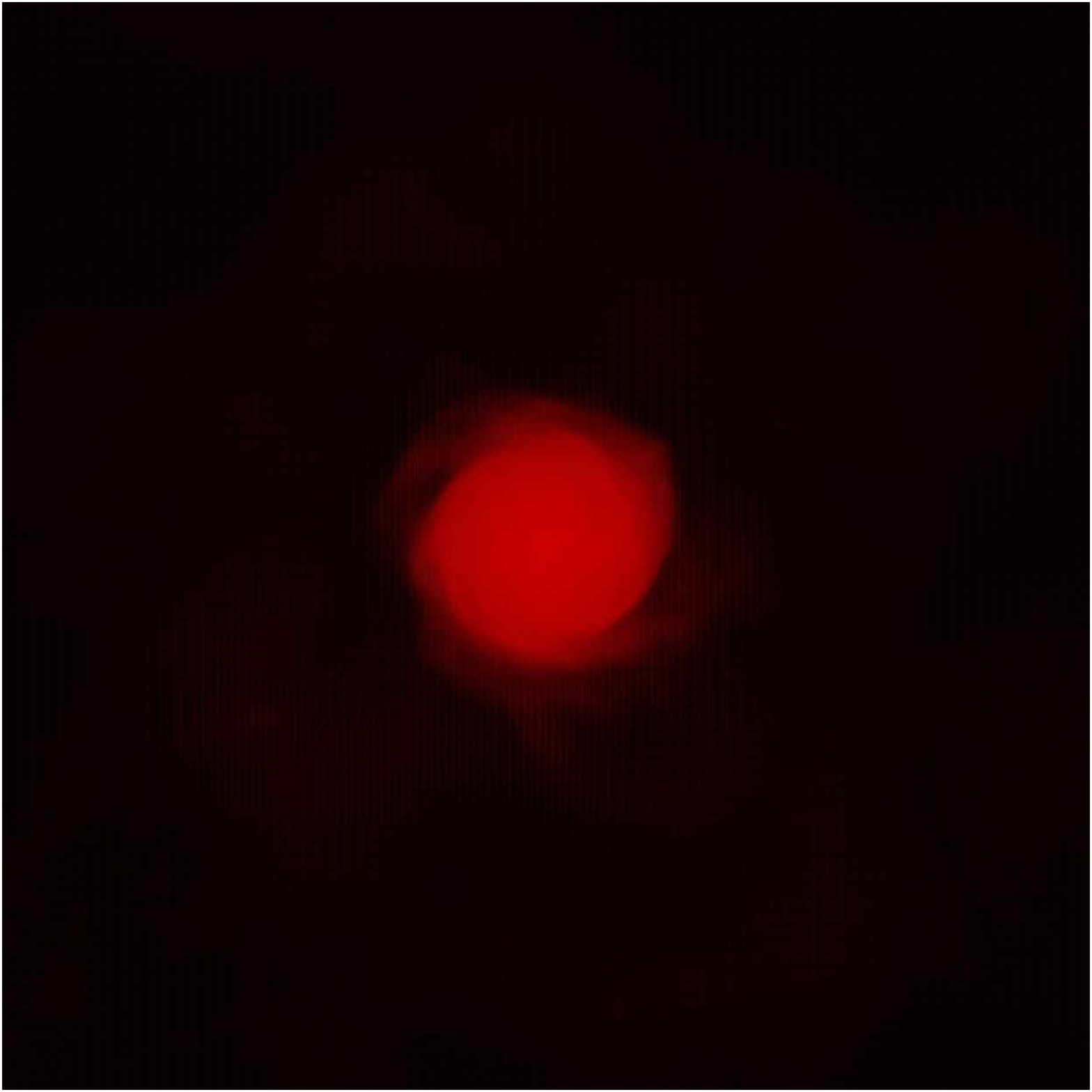}}
\subfigure{\includegraphics[scale=0.1]{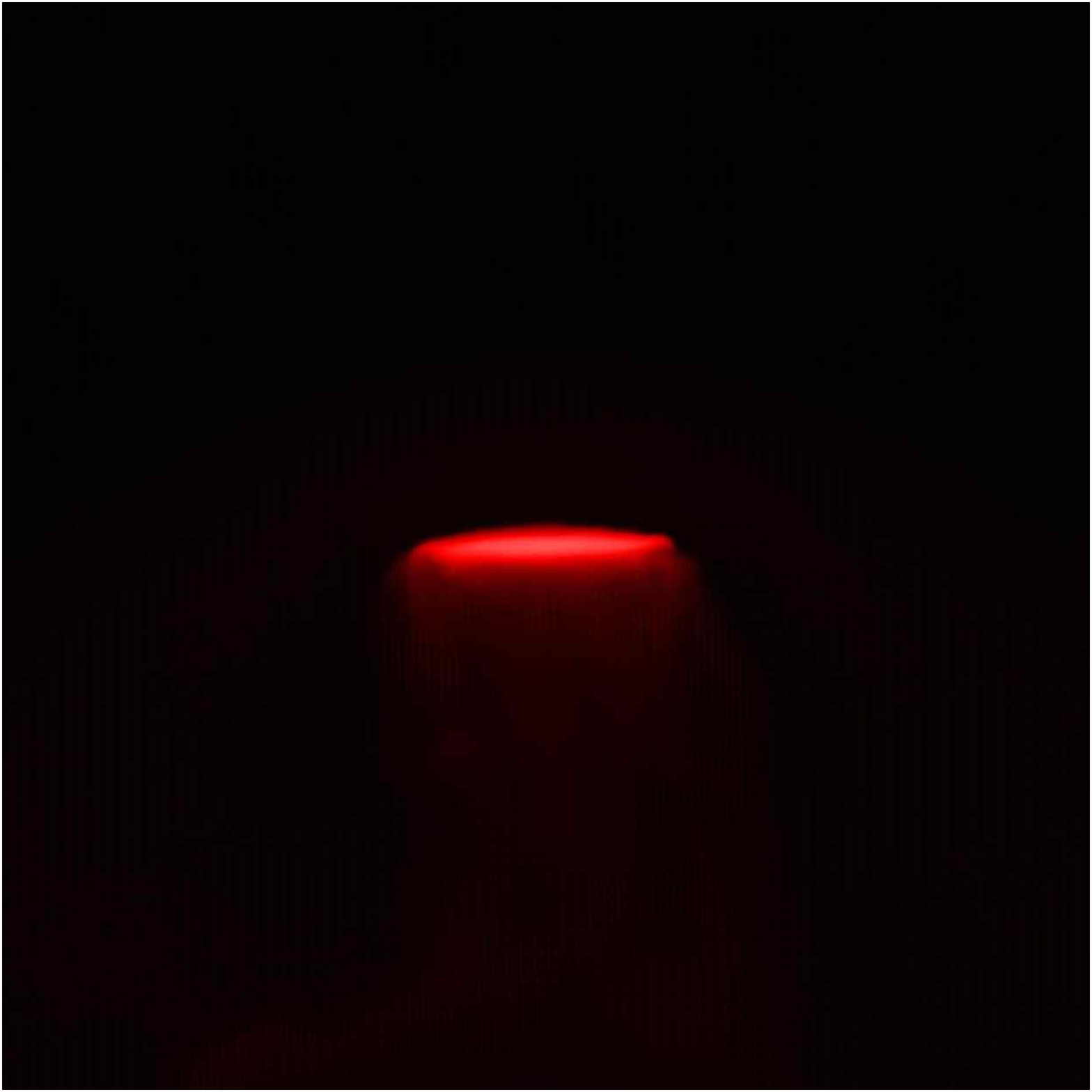}}\\
\caption{A face-on and edge-on view of gas surface density simulation PHNCW, at 40, 250, 500, and 750 Myr after the wind has hit the galaxy.  Note the continued smooth distribution of gas both in the disk and in the stripped gas throughout the simulation.  Both the box size and color scale are the same as in Figure \ref{fig:rcshots}}
\label{fig:ncshots}
\end{figure}

In Figures \ref{fig:rcshots} and \ref{fig:ncshots} we show projections along the y- and z- axis towards PHRCW and PHNCW about 40 Myr, 250 Myr, 500 Myr, and 750 Myr after the wind has hit the galaxy.  The projections are shaded to reflect surface density of the gas with a logarithmic stretch, and are 103.7 kpc across.  There are a number of important differences in the morphology of the gas disks that are reflected in our quantitative comparisons.  First, the gas in the galaxy with radiative cooling (PHRCW) forms small clouds with radii of about 100 pc near the center of the disk (within 7 kpc) that are of higher density than anywhere in the gas disk without cooling (PHNCW).  Only these central clouds attain molecular densities, even in the galaxy that is never affected by an ICM wind.  Further, at any radius of the disk in PHRCW, there is both higher and lower density gas than in the disk of PHNCW.  

This density structure has an important impact on the way that gas is stripped from the disk.  First, it allows ICM gas to create holes in the disk, which leads to ablation of material from a wide range of radii, rather than just at the outer edge of the disk, as in the no-cooling run (PHNCW).  As we will show in more detail below, this allows the gas to be stripped more quickly (see also Quilis et al 2000).  In addition, the gas that is stripped tends to stream directly behind the disk, rather than flaring to the side, and therefore creates a narrower wake.  We will examine the wake in more detail in a future paper.

The density fluctuations in the disk in the PHRCW galaxy create a ring of higher density gas from about 10 kpc to 16 kpc.  This gas never collapses to the highest densities.  In fact, in the PHRCNW galaxy, gas with $\rho > 10^{22}$ g cm$^{-3}$ only persists for $\sim$100 Myr in the middle of the simulation.  Although the vast majority of gas in this ring is always of a low enough density for direct stripping to occur, allowing the simulation to run until this ring had fully fragmented might result in a different stripping scenario.  It would also allow gas more time to drift towards the center of the galactic potential, making it more difficult to strip.  
However, because little molecular gas is found outside the center of galaxies in both these simulations and in observations (e.g. Leroy et al. 2008), we don't expect the ring structure to make much difference in how much gas can be stripped.

Although in this paper we will not discuss the stripped gas in any detail, it is notable that the stripped gas in PHRCW has more density substructure than that of PHNCW.  Higher density gas leaves the disk in PHRCW than in PHNCW (although not gas with molecular densities), and stripped clumps survive for at least 100 kpc.  

\subsection{Gas Mass Loss}

\begin{figure*}
\includegraphics[scale=1]{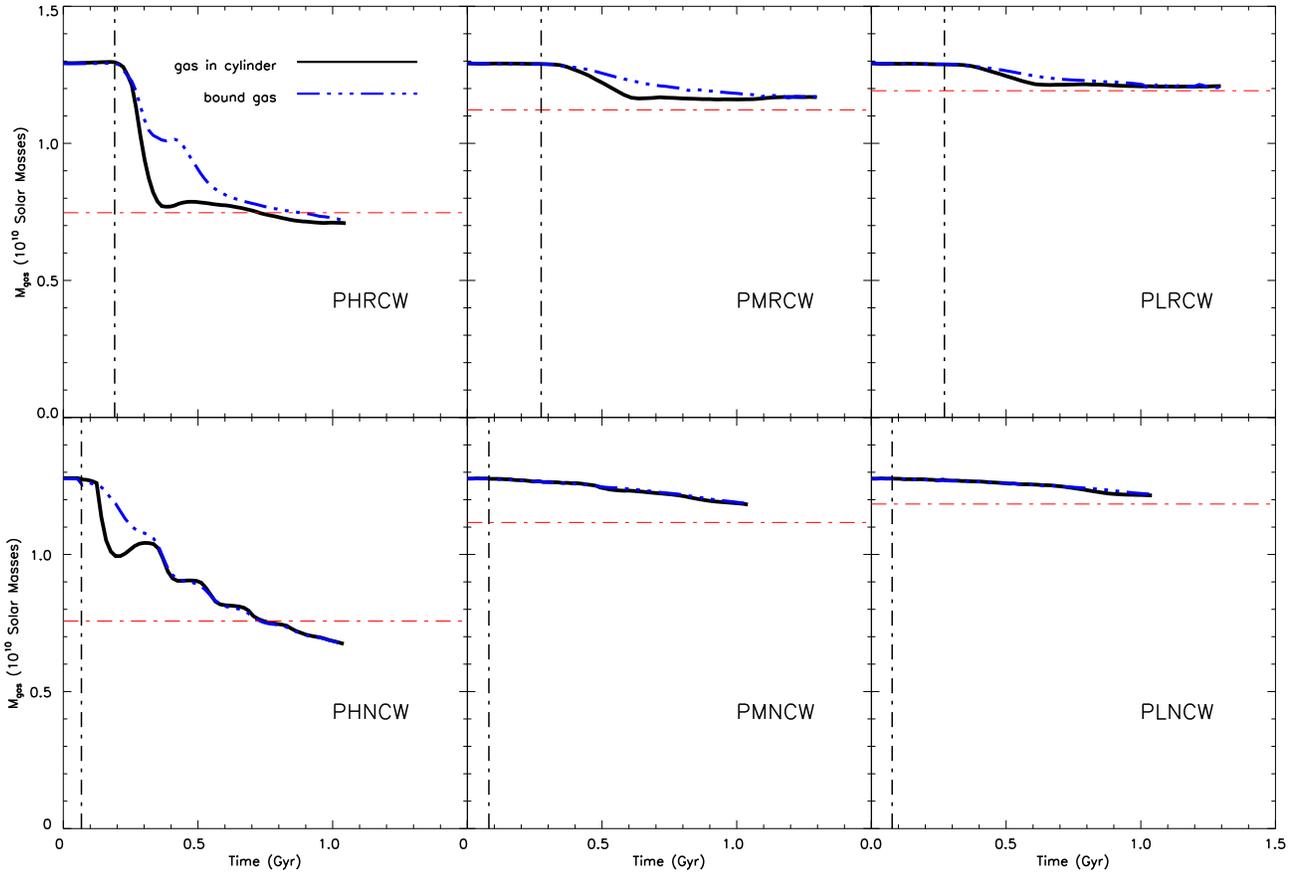}
\caption{The mass in a cylinder with radius 27 kpc and height 10 kpc centered on the stellar disk center of the galaxy, and the mass gravitationally bound to the galaxy plotted against time for the six cases effected with an ICM wind.  The columns are PH, PM, and PL.  The top row includes radiative cooling and the bottom row has no cooling.  The dash-dotted vertical line marks when the ICM wind hits each galaxy.  The dash-dotted horizontal line denotes the analytic prediction for the gas mass after ram pressure pushing.}
\label{fig:masses}
\end{figure*}

We begin our quantitative analysis of the impact of cooling by comparing the amount of gas mass that is lost in the cases with and without cooling (Figure \ref{fig:masses}).   In general, the amount of gas mass lost by the galaxies that include cooling is about the same as the gas mass lost by the galaxies that do not include cooling.  However, the timescale for stripping is much shorter for the radiatively cooled galaxies.  These differences are illustrated in Table \ref{tbl-strip}.  Note that the last column in this table is the time from when the wind first hits the galaxy to when the galaxy has lost 85\% of the total amount of gas lost.

In the highest ram pressure runs (PHNCW and PHRCW), after the wind has been hitting the galaxy for 800 Myr, both the galaxies with and without cooling have lost similar amounts of mass:  $5.845 \times 10^9$ M$_\odot$ and $5.86 \times 10^9$ M$_\odot$ in the cooling and no cooling cases, respectively.  We can compare these values to an analytic prediction for how much mass the galaxies should lose based on the amount of gas mass outside the analytic Gunn \& Gott (1972) stripping radius measured immediately before the wind hits the galaxy.  We find that the predictions are similar, although smaller, than the measured values:  the galaxy with cooling should lose $5.46 \times 10^9$ M$_\odot$ and the galaxy without cooling should lose $5.22 \times 10^9$ M$_\odot$.  In the case with cooling, almost the entirety (86\%) of this mass loss occurs within 150 Myr, while the same amount of mass loss takes about 650 Myr in the PHNCW case.  Also, both of these galaxies have a fallback event (gas that is initially stripped from the disk region, although still gravitationally bound, falling back onto the disk) about 250 Myr after the wind hits the galaxy, although in PHNCW, the fallback is more drastic and seems to be followed by a few minor fallback events.  Although the PHRCW case seems to stop losing gas despite the continued ICM wind, losing less than $5 \times 10^8$ M$_{\odot}$ in the final 300 Myr we follow the galaxy, the PHNCW case does not slow its gas loss by nearly as much, losing twice as much gas in the same amount of time.  This difference is fit well by the difference in radius between the two cases and the dependence of Kelvin-Helmholtz stripping on radius squared.

In PMRCW, our run with the middle ram pressure value, after 900 Myr of the wind hitting the galaxy, it has lost about $1.3 \times 10^9$ M$_{\odot}$ of gas mass.  In fact, it loses almost this entire amount of gas (95\%) within 340 Myr of the wind first hitting the galaxy.  On the other hand, in PMNCW, the galaxy has only lost $9.5 \times 10^8$ M$_\odot$ in the 900 Myr since the ICM wind has hit the galaxy.  Unlike the highest ram pressure case, the analytic prediction for the amount of gas loss is larger than the measured amount:  $1.7 \times 10^9$ M$_\odot$ for PMRCW and  $1.6 \times 10^9$ M$_\odot$ for PMNCW.  Again, the flatness of the PMNCW gas mass curve and the steeper grade of the PMRCW gas mass curve may be attributed to Kelvin-Helmholtz stripping effecting only the galaxy without cooling.  Also, there does not seem to be any steep drop in the PMNCW galaxy's gas that is the hallmark of ram pressure pushing.  Even more than in the highest ram pressure case, the galaxy with cooling seems to be only instantaneously stripped, without much longer-term Kelvin-Helmholtz stripping.  

The lowest ram pressure runs (PLRCW and PLNCW) are very similar to the middle ram pressure runs, except that only $8.5 \times 10^8$ and  $5.8 \times 10^8 \msun$ of gas are lost in the galaxies with and without cooling, respectively.  Like the middle ram pressure case, the analytic prediction for the amount of gas loss is more than the amount measured, $1.0 \times 10^9$ (PLRCW) and  $9.3 \times 10^8$ (PLNCW) M$_{\odot}$.  Even when the ram pressure is only $6.4 \times 10^{-13}$ dynes cm$^{-2}$, the galaxy with radiative cooling appears to lose gas through ram pressure pushing.  

Although similar amounts of gas mass are lost whether or not the galaxy is allowed to radiatively cool, there are a few trends.  First, in the middle and low ram pressure cases, the amount of gas lost by the galaxy with radiative cooling is more than that lost by the galaxy without radiative cooling.  Although this qualitatively agrees with the analytic predictions for the amount of gas mass loss indicating that the radiatively cooled galaxy will lose more gas, the difference in measured mass loss between the RC and NC cases is greater than the difference in the analytic prediction.  Thus, it may be that for lower ram pressures, a multiphase ISM is easier to strip.  Also, galaxies that radiatively cool lose their gas more quickly.  There is slightly more mass in total outside the stripping radius in the radiatively cooling galaxies, and there is more mass at higher density.  A plausible reason for the quick stripping of gas mass, then, is that a lumpy gas distribution is more quickly and easily stripped than gas with a smoother profile.  Later continuous stripping seems to mainly affect the galaxies without cooling, perhaps because their larger cohesive gas disk is more affected by the Kelvin-Helmholtz instability.  

\begin{table}
\begin{center}
\caption{Stripping Data\label{tbl-strip}}
\begin{tabular}{c | ccccc}
\tableline
Runs & M$_{lost}$ & M$_{G\&G}$ & r$_{strip}$ & r$_{G\&G}$ & t$_{85\%}$\\
\tableline
PHRCW & $5.845 \times 10^{9}$& $ 5.46 \times 10^9$ & 11& 11.46 & 148  \\
PHNCW &  $5.86 \times 10^{9}$ & $ 5.22 \times 10^9$ & 7 & 11.46 &  625 \\
\tableline
PMRCW & $1.3 \times 10^{9}$& $1.7 \times 10^9$ & 11 & 16.8 & 288 \\
PMNCW & $9.5 \times 10^{8}$ & $1.6 \times 10^9$ & 19.5 & 16.8 & 810 \\
\tableline
PLRCW & $8.5 \times 10^{8}$& $1 \times 10^9$ & 15 & 18.22 & 316 \\
PLNCW & $5.8 \times 10^{8}$ & $9.3 \times 10^8$ & 20? & 18.22 & 754 \\
\tableline
\tableline
\end{tabular}
\end{center}
\end{table}

\subsection{Stripping Radius}

\begin{figure*}
\includegraphics{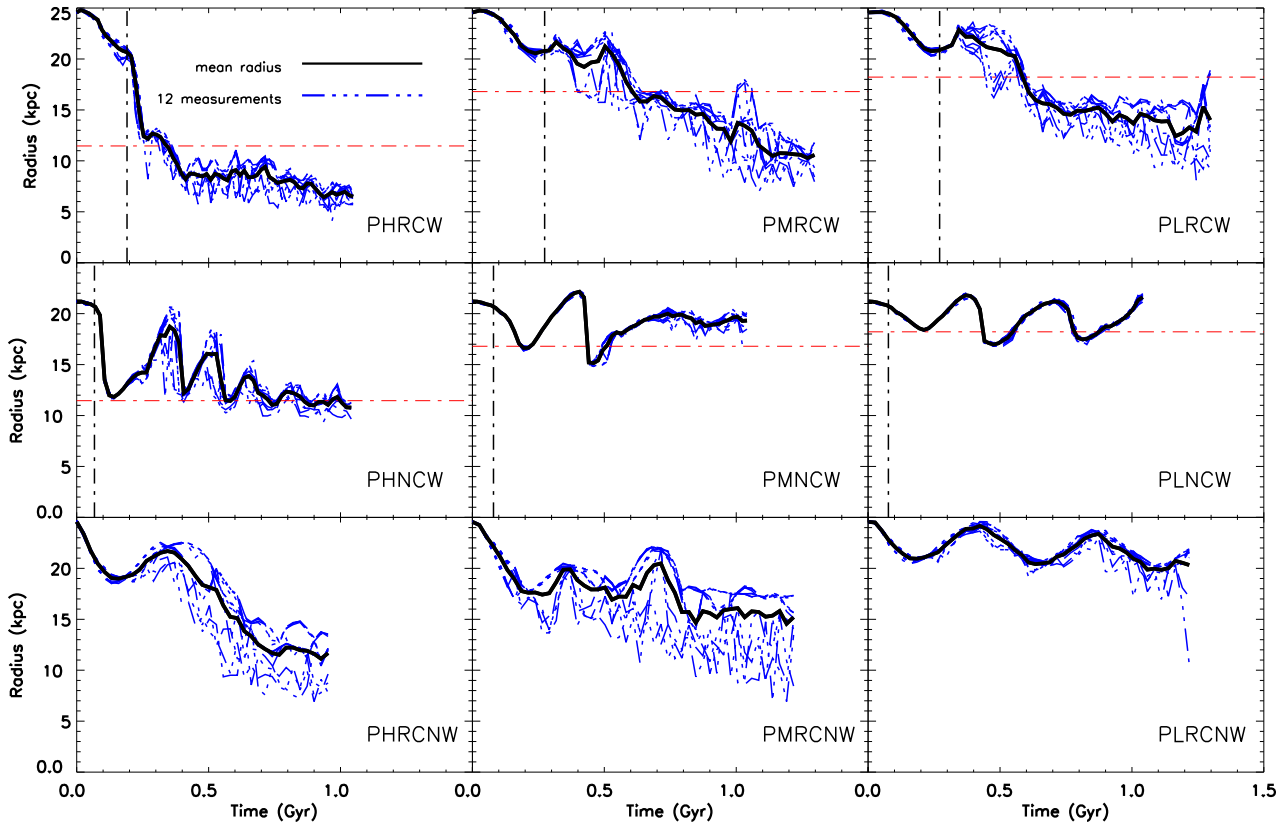}
\caption{The radius of gas with a density greater than 10$^{-26}$ $\rho$ cm$^{-3}$.  The columns are PH, PM, and PL.  The top row includes radiative cooling, the middle row has no cooling, and the bottom row has cooling but no ICM wind.  The dash-dotted vertical line marks when the ICM wind hits each galaxy.  The dash-dotted horizontal line denotes the analytic prediction for the stripping radius based on the initial gas distribution.}
\label{fig:radii}
\end{figure*}

In general, the galaxies that include radiative cooling are stripped to a smaller radius than predicted using the Gunn \& Gott (1972) analytic equation, and end at a smaller radius than stripped galaxies without cooling and at a smaller radius than galaxies with radiative cooling that are not affected by an ICM wind (Figure \ref{fig:radii}).  Again, see Table \ref{tbl-strip} for the stripping radius values. 

In the highest ram pressure cases, the analytic stripping radius is 11.5 kpc, very similar to the stripping radius of the galaxy in PHNCW.  The oscillations in the radius correspond to fallback episodes in Figure \ref{fig:masses}.  The final radius of the PHRCW case is about 7 kpc.  The final radius of the PHRCNW galaxy is just above 11 kpc.  This means that including cooling without a wind causes galactic gas to spiral towards the center of the disk, most likely because the entrainment of nonrotating ICM into the disk creates drag and angular momentum loss.  The entrainment of gas into the disk means that the average gas density of the resulting disk would be higher than before, making it more difficult to strip.  In fact, comparing the amount of gas at different densities in PHRCW and PHNCW about 15 Myr before the ICM wind hits the galaxy, we find that PHRCW has more high density gas outside 10 kpc and a higher fraction of the gas in the disk region is high density than in PHNCW.  Therefore, the holes in the disk must allow for easier stripping down to smaller radii.  The holes may also create more drag against surviving gas, allowing it to spiral towards the center of the disk more quickly.  

In the middle ram pressure cases, the analytic stripping radius is 16.8 kpc.  The final radius of the galaxy in PMNCW is between 19 and 20 kpc.  However, the initial radius drop is to about 17 kpc, while a static galaxy without cooling, only undergoing epicyclic variations, has a smallest radius of about 19 kpc.  The final average radius of the galaxy in PMRCNW is steady at 15 kpc, although it is very asymmetric.  The galaxy in PMRCW is stripped down to a radius of 11 kpc.  Again, a mixture of stripping and cooling results in the smallest radius.

In the lowest ram pressure cases, the analytic stripping radius is 18.2 kpc.  It is difficult to say whether the galaxy in PLNCW is stripped to a smaller radius because of the large oscillation in radius over time.  PLRCW is stripped to about 15 kpc.  In this case more than in the previous two, it is clear that it must be the holes punched by the ICM wind in the gas disk that allow the gas to be stripped to a small galactic radius.  Neither PLNCW nor PLRCNW have much change in the radius of the gas disk, so it must be a combination of the ICM wind and the clumpiness of the gas in the disk that result in the lowest gas radius.  

By examining the stripping radius of gas in these galaxies, we find that allowing radiative cooling results in a smaller radius when hit by an ICM wind.  Again, the most likely explanation is that the clumpiness of the disk gas allows for stripping to a smaller radius.

\subsection{Densest Gas}

As gas in the galaxy cools, it becomes more dense, eventually attaining densities greater than 1000 cm$^{-3}$, certainly in the range of molecular gas.  Although with a resolution of 40 pc, we only marginally resolve giant molecular clouds, we can begin to answer whether ram pressure has any effect on the densest gas in a galaxy (and hence on star formation).  

\begin{figure*}
\includegraphics{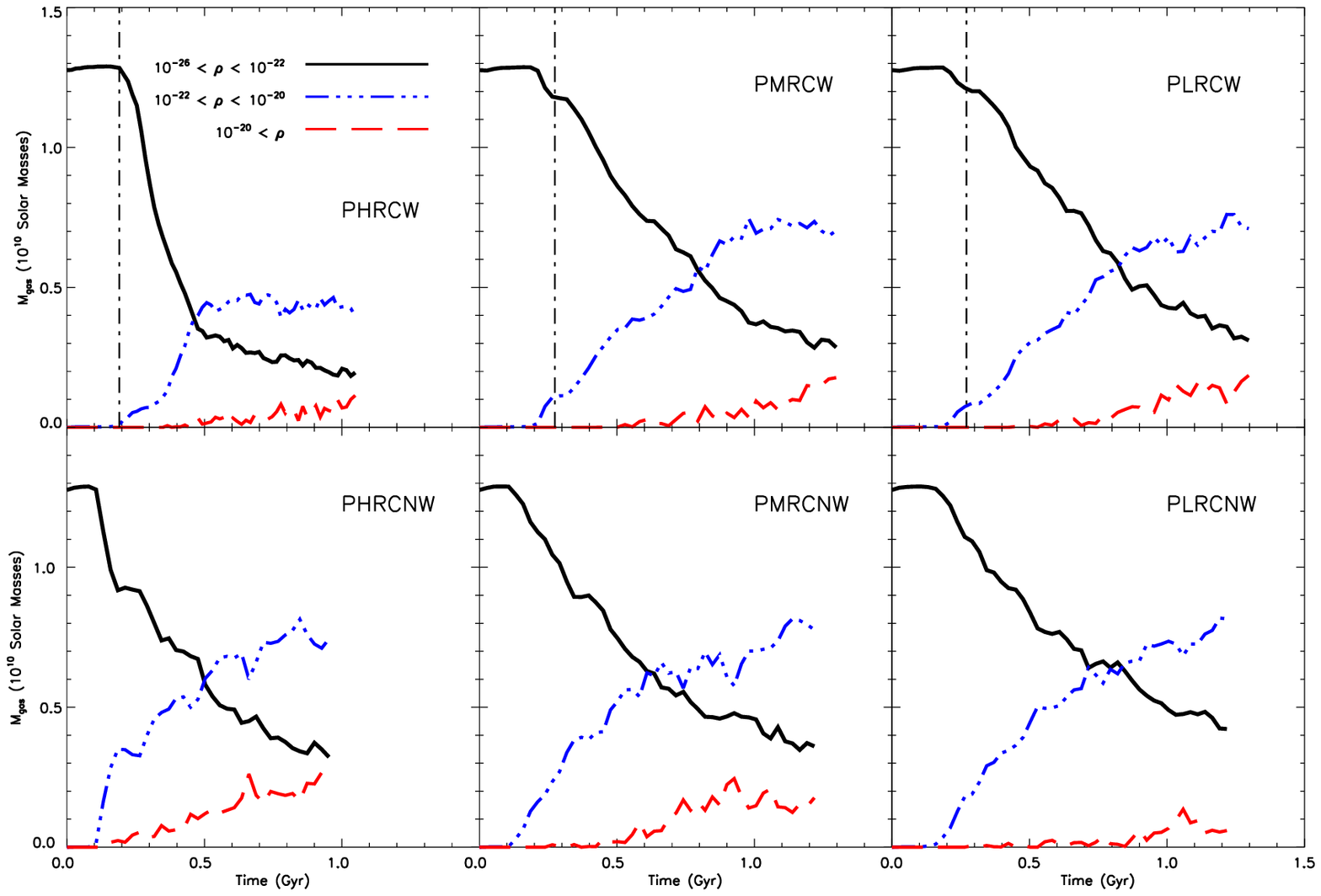}
\caption{The amount of mass at different densities contained within the cylinder surrounding the galaxy.  The columns are PH, PM, and PL.  The top row includes radiative cooling and the bottom row has cooling and no wind.  The dash-dotted vertical line marks when the ICM wind hits each galaxy.}
\label{fig:mass_rho}
\end{figure*}

When comparing the three cases with radiative cooling and no ICM wind (bottom panels in Figure \ref{fig:mass_rho}), it is clear that a higher pressure ICM results in a faster formation of high density gas and the formation of more high density gas.  However, as we will see, including an ICM wind can either strip the surrounding lower density gas that would accrete onto clouds, or create more high density gas when the wind pressure is low.

\begin{figure*}
\includegraphics{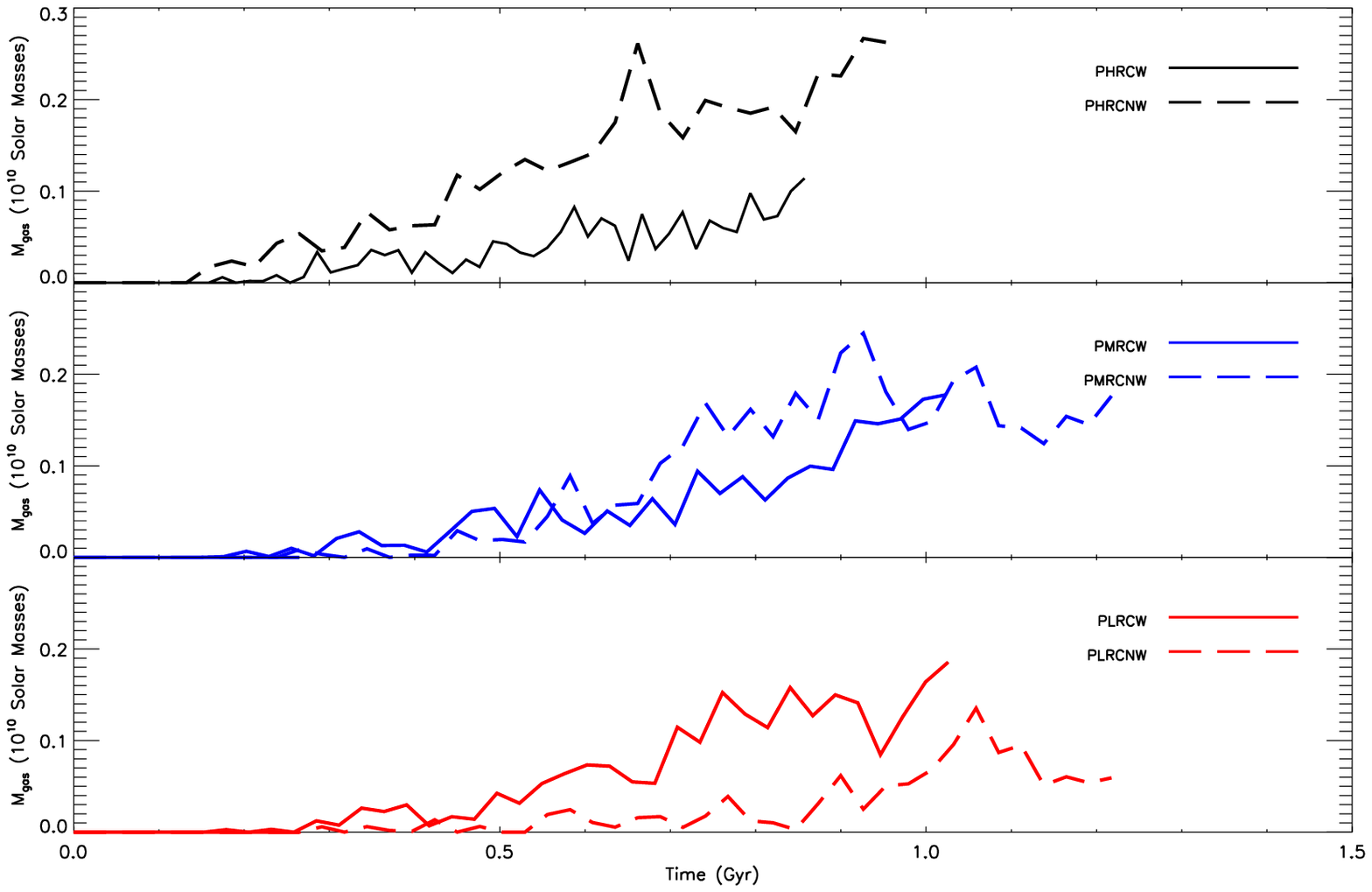}
\caption{This focuses on the most dense gas in each of the radiatively cooled cases.  The runs with an ICM wind have been shifted in time by the amount of time before the wind hits them (190 Myr, 274 Myr, and 271 Myr, respectively).  High ram pressure results in less high density gas, while low ram pressure increases the amount of high density gas.}
\label{fig:mass_rhoh}
\end{figure*}

We find that with the highest ram pressure of $6.4 \times 10^{-12}$ dynes cm$^{-2}$, the amount of higher density gas (above $10^{-22}$ g cm$^{-3}$) is less than if the galaxy were to radiatively cool without an ICM wind (see Figure \ref{fig:mass_rho}).  This is true even when (as in Figure \ref{fig:mass_rhoh}), we shift the PHRCW line showing how much gas has cooled as a function of time by 200 Myr in order to account for the fact that the galaxy in PHRCW has been in the higher pressure ICM for almost 200 Myr less than the galaxy in PHRCNW.  When the wind hits the galaxy, less dense gas throughout the disk is accelerated out of the galaxy, leaving less nearby lower density gas that would later either be accreted onto nearby clouds or cool and collapse into new clouds.  However, gas with densities above $10^{-22}$ g cm$^{-3}$ is never seen outside of the disk (always within 1 kpc of the disk plane), so molecular clouds are not being stripped in their entirety.  Examining the snapshots indicates that higher density clouds do leave the disk, leading us to hypothesize that dense clouds can be ablated until they are of just low enough column density to be stripped from the disk.   

The scenario seems less clear in the medium ram pressure case ($1 \times 10^{-12}$ dynes cm$^{-2}$).  Although the amount of higher density gas increases more slowly in PMRCW than in PMRCNW, both end their runs with similar amounts of high density gas (see Figure~\ref{fig:mass_rhoh}).  

In the lowest ram pressure case ($6.4 \times 10^{-13}$ dynes cm$^{-2}$), the galaxy that is hit by the ICM wind has more of the highest density gas (above $10^{-20}$ g cm$^{-3}$) than the galaxy that evolves in the post-shock ICM.  In this case, the added ram pressure from the wind appears to compress the gas, leading to more higher density gas.

In summary, molecular clouds do not get stripped, even if the ram pressure is $6.4 \times 10^{-12}$ dynes cm$^{-2}$.  They do not grow within a high ram pressure ICM wind, however, because lower density gas is accelerated out of even the inner disk (inside the stripping radius of 7 kpc).  However, with low ram pressure, $6.4 \times 10^{-13}$ dynes cm$^{-2}$, more high density gas is formed, likely because the ram pressure, while not enough to strip much gas from the disk, is enough to compress gas to a higher density.

\section{Discussion}\label{sec-discussion}

\subsection{Comparison to Observations}

A galaxy is considered HI deficient if its deficiency value is 0.3 or greater.  Deficiency is defined as DEF = $<$ log X $>$$_{\rm T,D}$ - log X$_{\rm obs}$ (Giovanelli \& Haynes 1985), where X$_{\rm obs}$ is the observed HI mass of the cluster galaxy and $<$ log X $>$$_{\rm T,D}$ is averaged over a sample of field galaxies of morphological type T and optical diameter D.  Given this definition, the highest ram pressure case is almost but not quite HI deficient.  However, gas is truncated to less than half the original gas radius in the largest ram pressure case, and is down to less than 75\% in the the lowest ram pressure case.  If we were to roughly classify these galaxies using the scheme of Koopmann \& Kenney (2004), assuming that HII regions will follow our gas distribution, all three of our runs result in truncated disks.  

We can also compare the amount of high density gas we find in our galaxies to the amount of molecular gas observed using CO.  Kenney \& Young (1986, 1989) found that HI deficient galaxies in Virgo have undepleted CO, and Nakanishi et al. (2006) found that the molecular fraction in three galaxies near the center of Virgo is high.  We find that in our high ram pressure run, the fraction of high density (i.e. molecular) gas actually remains similar to the case with no wind.    On the other hand, in the low ram pressure case, the fraction of high density gas does increase as the ICM wind compresses the ISM.  This case better matches the observations of high molecular gas fractions (Nakanishi et al. 2006).  It is possible that galaxies will move through low ram pressure environments on their way to higher ram pressure, so the initial compression followed by strong stripping could result in HI depleted disks with unaffected, or increased, amounts of molecular gas.  However, we do note that these results are somewhat sensitive to the resolution and details of the cooling (discussed in more detail below).

We do not include star formation in our simulations, but we do have an enhanced amount of high density gas in our lowest ram pressure simulation, while our highest ram pressure simulation has less high density gas in comparison to their respective no-wind comparison runs.  If we assume that higher density gas corresponds to a higher rate of star formation, the lowest ram pressure case would result in an enhanced star formation rate, while the highest ram pressure case would result in a lower star formation rate.  Placing this into a cluster environment, we would expect to find enhanced star formation at or outside the virial radius in the lower density ICM, and post-star-forming galaxies closer to the cluster center.  In Coma, Poggianti et al (2004) found that post-starburst or post-star-forming galaxies were found at the edge of ICM substructure, which seems to follow our claim that starbursts will occur outside of the densest ICM, while high ram pressure will quench star formation.

We can also use morphological cues to determine how closely these simulations match observations.  A perfect example of a galaxy showing both the strength and weakness of our simulations that include a multiphase medium is the Virgo spiral NGC 4402 (Crowl et al 2005).  The galaxy has upturned edges in it's dust and gas distribution, similar to the earliest projection of our galaxy without cooling.  However, similar to our galaxy with radiative cooling, there is evidence that two dense clouds have survived longer than the lower density ISM and are being ablated by the wind.    

\subsection{The Evolving ISM}

A molecular cloud may survive between a few and a few tens of Myr (Blitz \& Shu 1980; Larson 2003; Hartmann 2003).  Because we do not include star formation, our dense clouds of gas do not evolve into less dense pockets of gas that could be stripped.  In the highest ram pressure case, holes are punched around the disk (as close as 6 kpc to the disk center) between 6 - 11 Myr after the ICM wind first hits the galaxy, which indicates that as long as gas stays low density for 10 Myr it can be stripped by the wind.  If our dense regions of gas could turn into very low density regions through star formation ($\rho$ $\le$ 10$^{-23}$ g cm$^{-3}$), our galaxies that include radiative cooling could be entirely stripped of gas in the highest ram pressure case.  In the lowest ram pressure case, no gas appears to be stripped from the central region of the gas disk.  

\subsection{Magnetic Fields and Cosmic Rays}

Our simulation does not include magnetic fields or cosmic rays.  In general in the Milky Way, the magnetic field is measured to be about 2-3 microGauss (Men et al 2008).  A field of this strength would not have enough magnetic pressure ($2-5 \times 10^{-13}$ dynes cm$^{-2}$)  to significantly lessen the direct impact of ram pressure on the galactic gas.   However, there are measurements of the magnetic field in molecular clouds (Crutcher 1991 and references therein) of up to $10^3$ microGauss, which can certainly withstand ram pressures of order 10$^{-12}$ dynes cm$^{-2}$.  These magnetic fields could help protect molecular clouds from being ablated by the ICM wind, so could help them survive.  Also, a magnetic field could attach a molecular cloud to less dense gas, either anchoring the surrounding gas or helping to drag out the molecular cloud.  Using this simulation we cannot conjecture on the likely impact of magnetic fields on the total amount of gas stripped.

We also do not include cosmic rays in our simulation, which would add energy to the galactic gas and increase the vertical distribution (e.g. Joung \& Mac Low 2006).  It is also possible that the increase in gas pressure would lower the amount of entrainment into the disk from the lower pressure ICM, which would in turn lessen the number of very low density regions and make the disk more cohesive.  This would make stripping more difficult because fewer holes could be punched through the disk by the ICM wind.  However, the energy input from cosmic rays would slow high density gas formation, possibly allowing for more extended stripping if the gas is not locked in very dense clouds.  As with magnetic fields, the addition of cosmic rays into our simulation does not have an easily predicted effect.

\begin{figure*}
\includegraphics{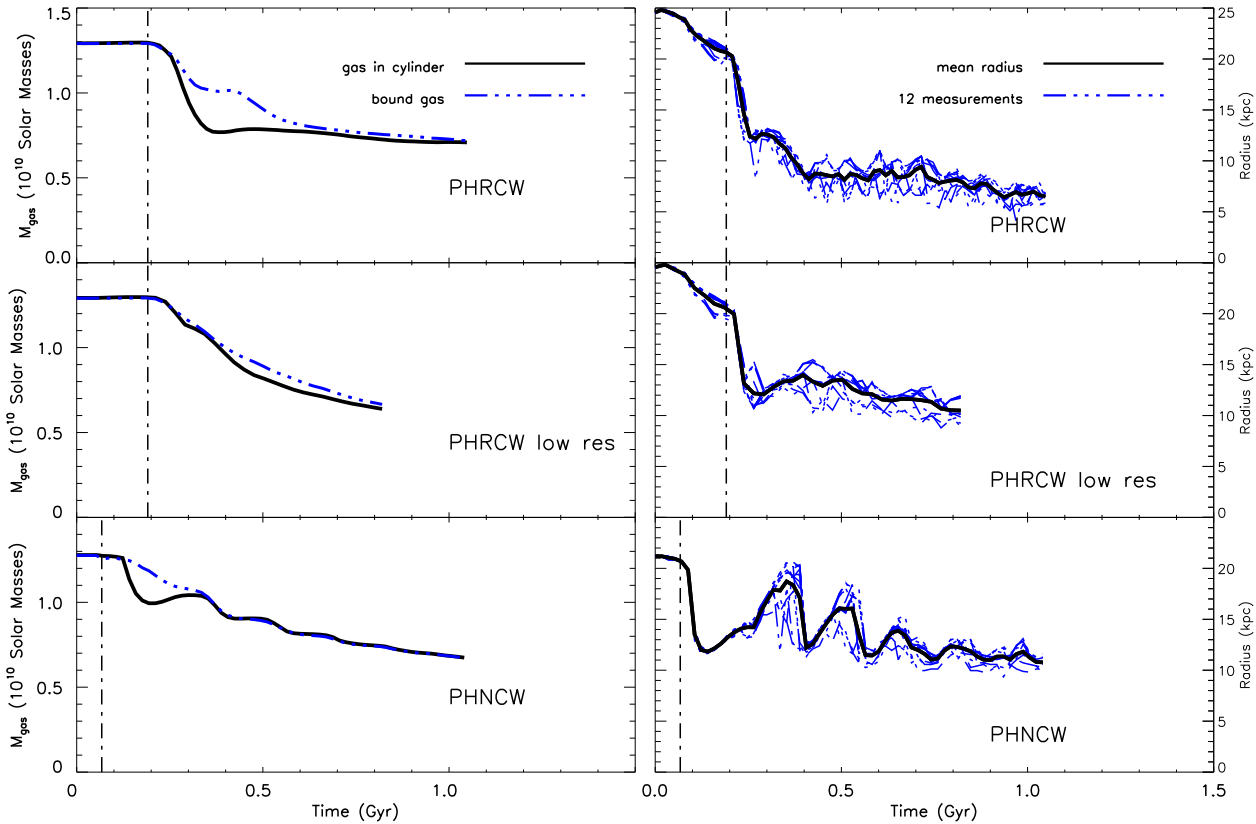}
\caption{The gas mass and gas radius over time for PHRCW, PHRCW low resolution, and PHNCW.  This figure shows how the low resolution run with radiative cooling is a cross between PHRCW and PHNCW, because the gas density is smoother than PHRCW, but less smooth than PHNCW.  See \S \ref{sec-res} for discussion.}
\label{fig:resapp}
\end{figure*}

\subsection{Resolution}\label{sec-res}

As has been discussed in detail in previous works (e.g. Tonnesen et al. 2007 and Roediger \& Br\"uggen 2006), resolution does not have a large effect on our runs without radiative cooling.  The runs without radiative cooling that we discuss in this paper have a cell size of 80 pc, and we ran a comparison run with a cell size of 40 pc to check this.  We found very similar results in the amount of gas and the radius of the disk.  In addition, our results are similar to those in Roediger \& Br\"ggen (2006), where even lower resolution was used.

However, including radiative cooling leads to results that are very dependent on resolution.  The runs discussed in this paper have a cell size of 40 pc, which leads to gas collapsing to molecular densities.  For example, the surface density in the center of the disk immediately before the wind hits the galaxy is 200 M$_\odot$ kpc$^{-2}$, with values above 10 M$_\odot$ kpc$^{-2}$ for much of the inner disk (this surface density corresponds to the transition between HI and molecular gas observationally found in Bigiel et al 2008).

\begin{figure}
\includegraphics[scale=0.5]{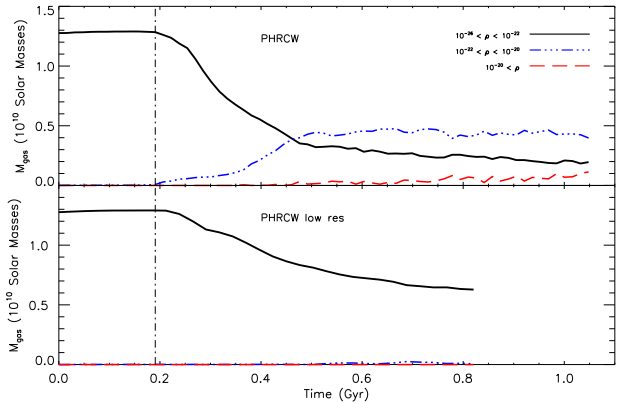}
\caption{The amount of mass at different densities contained within the cylinder surrounding the galaxy for the strong ICM wind case at high and low resolution.  The lower resolution run does not form the clouds of molecular density, so does not model the multiphase medium we attempt to study.  See \S \ref{sec-res} for discussion.}
\label{fig:resapp_rho}
\end{figure}

To examine the impact of resolution, we have run a comparison simulation with the same parameters as PHRCW, but with a cell size of 80 pc.  We find that gas density does increase, but the highest density gas never forms, and there is very little gas in the middle density range.  Because the lack of cloud formation results in a smoother gas distribution in the galaxy disk, the ICM wind does not have the same effect on this lower resolution disk as it does in the higher resolution cases.  Figure \ref{fig:resapp} shows that although the low resolution disk with cooling still gets stripped slightly more quickly than the no cooling case, it does not get stripped as quickly as the higher resolution case.  This occurs because the corresponding low density regions in the disk do not form, so the ICM wind does not punch holes throughout the disk.  The low resolution run also has much less fallback than PHNCW, but because very dense clouds are not formed (see Figure \ref{fig:resapp_rho}), continuous stripping occurs and gas loss does not cease after a few hundred Myr.

We also ran a pair of comparison runs between a galaxy with a cell size of 40 pc and one with a cell size of 20 pc.  Because of the computing time needed to run the high resolution simulation, we only had output for 160 Myr of evolution.  However, within this time, the maximum surface density of the higher resolution run is more than double that of the regular resolution run.  Further, after 160 Myr the amount of gas with densities above $10^{-20}$ g cm$^{-3}$ is already more than 25\% of the total galaxy gas mass and growing at a rate of 40 M$_\odot$ yr$^{-1}$, while gas with those densities has not yet formed in the regular resolution run (at 160 Myr, $\rho_{\rm max}$ $\geq$ 10$^{\rm -22}$ g cm$^{-3}$).  In fact, in the highest resolution run, $5 \times 10^8$ M$_{\odot}$ of gas with densities above $10^{-19}$  g cm$^{-3}$ ($n > 10^{5}$ cm$^{-3}$) has already formed by 160 Myr into the simulation, with no sign of slowing down much higher gas density formation.  A visual inspection shows that the cloud radius is set by the minimum cell size in the simulation.  We conclude that there are physical processes that we do not include in the simulation, such as star formation, magnetic fields, cosmic rays, and feedback from supernovae, that would all result in lower density gas than is found in our simulations, either by adding energy to the ISM or locking the ISM into place and hindering collapse.  In this case, higher resolution is not necessarily better, because our omissions may result in unphysical gas densities.  In the regular resolution run, the larger cells balance, to a certain extent, the smaller-scale physics we omit.

\subsection{Radiative Cooling Floor}\label{sec-cool}

As mentioned in \S \ref{sec-sim}, we include radiative cooling using the Sarazin \& White (1987) cooling curve.  We allow cooling to 8,000 K, as this results in gas with neutral hydrogen temperatures without overcooling a large fraction of our gas.  With this cutoff, we find that we form clouds with densities and sizes typical of molecular clouds, although we make no attempt to reproduce the internal structure of the clouds.  For a more detailed discussion of the density distribution in the ISM for similar simulations, but without a wind, see Tasker \& Bryan (2006, 2008).   

To examine the impact of our adopted cooling floor, we perform a run with no wind and cooling to 300 K which shows that after 800 Myr the density distribution of disk gas was nearly identical to the case with cooling to 8000 K, although the case with cooling to 300 K reached the distribution more quickly and leveled out.  The highest density gas is less than a factor of three more dense in the 300 K case than in the 8000 K case.   

\begin{figure*}
\includegraphics{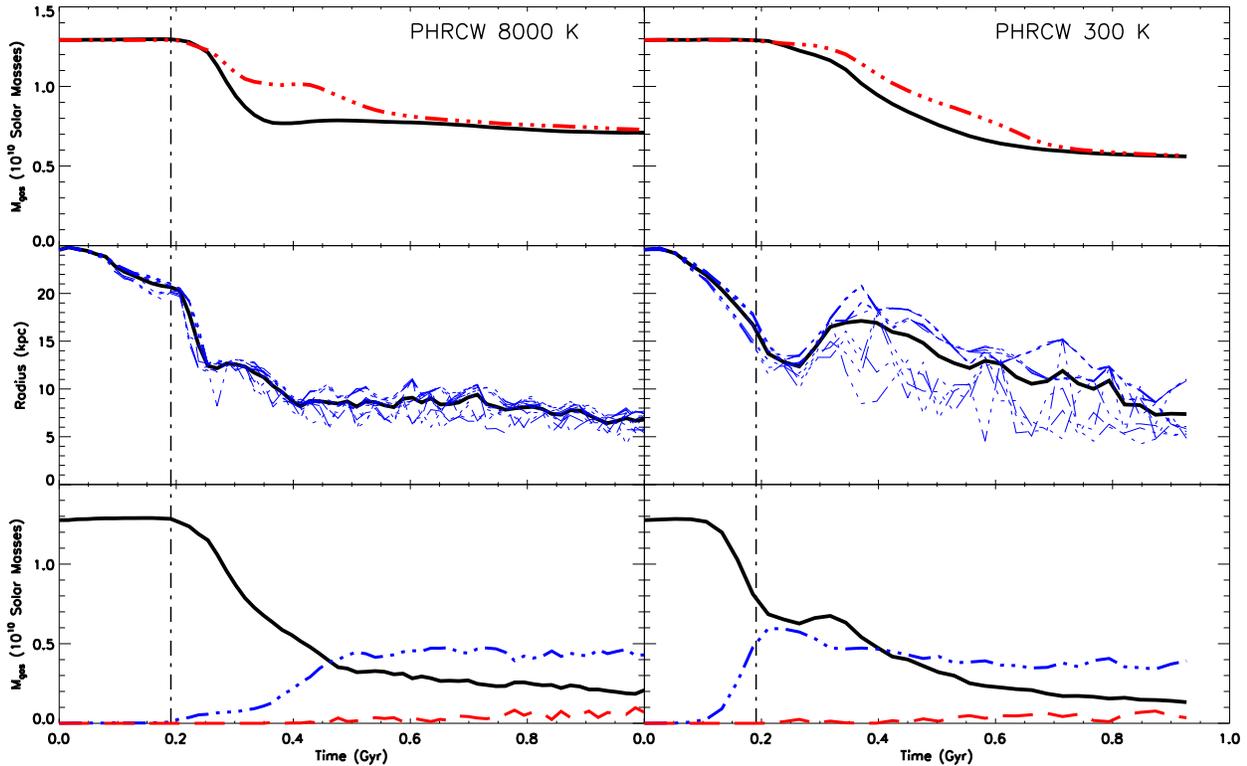}
\caption{Comparison between PHRCW and the case with cooling to 300 K.  In both cases, the ICM wind hits the galaxies at 190 Myr.  Note how extended the stripping is in the 300 K cooling case, and note that more gas is stripped.  These facts are linked with the larger gas radius for much of the simulation.  See \S \ref{sec-cool} for discussion.}
\label{fig:coolcomp}
\end{figure*}

\begin{figure*}
\centerline{
\subfigure{\includegraphics[scale=0.35]{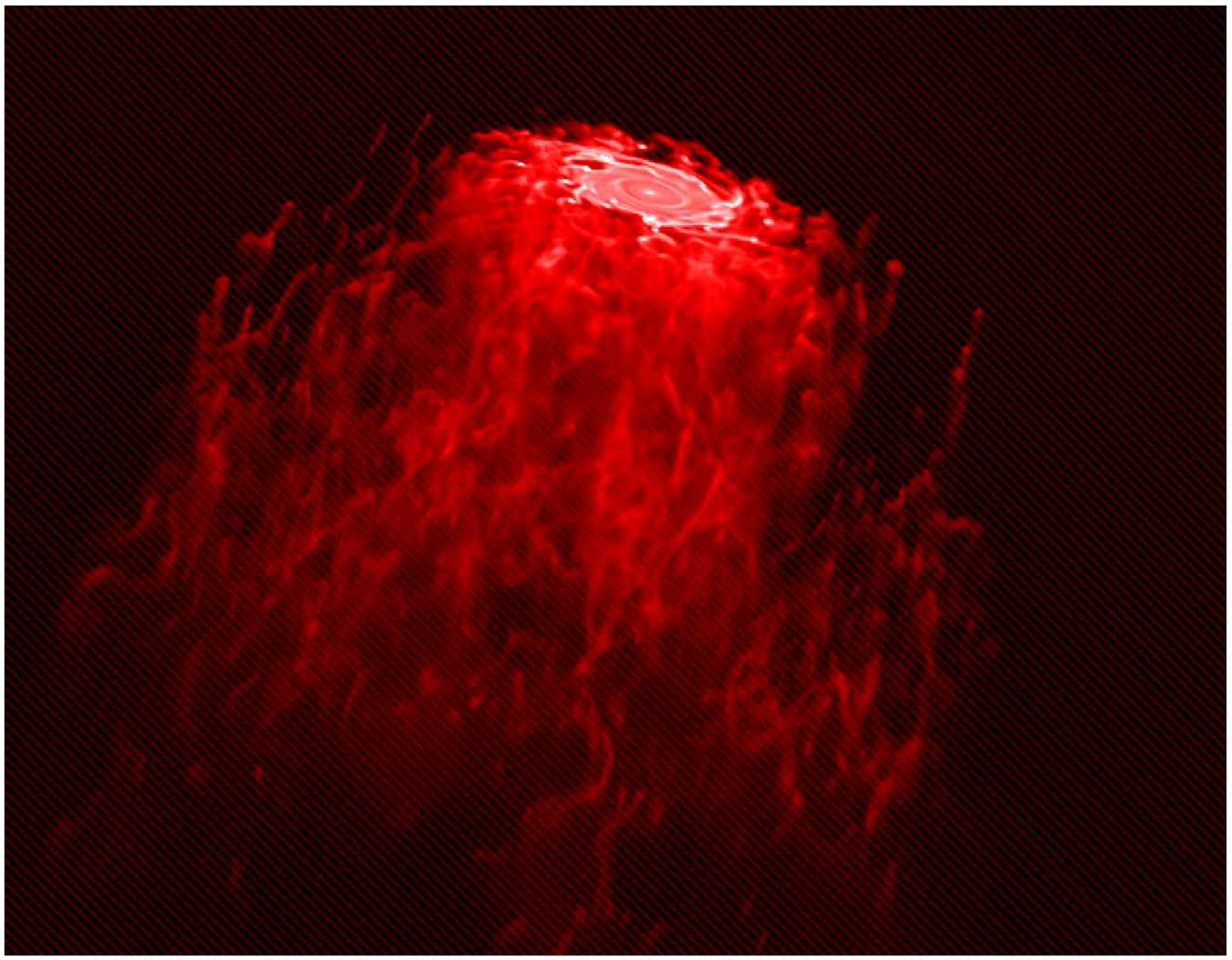}}
\subfigure{\includegraphics[scale=0.35]{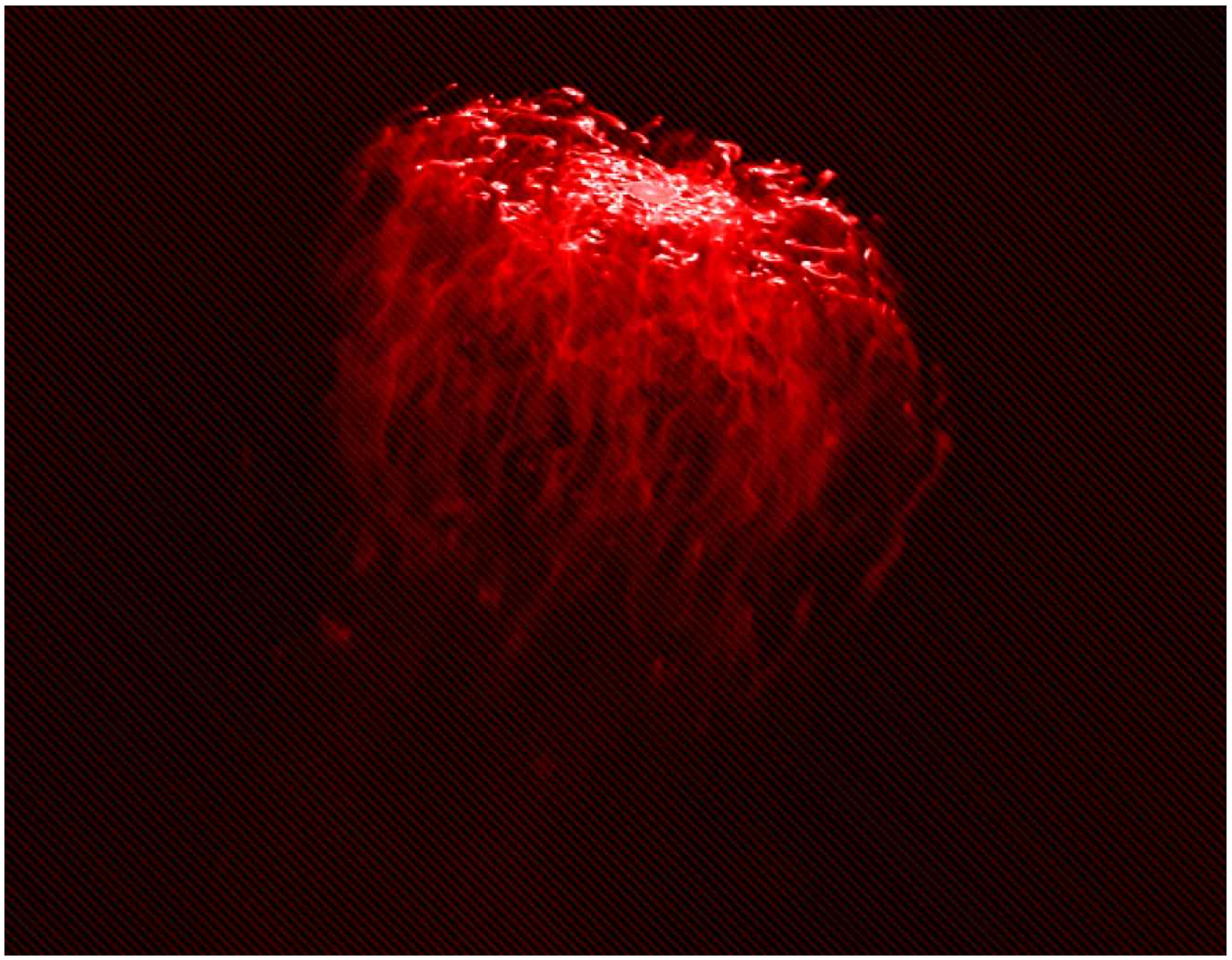}} }
\caption{A  tilted view of the high ram pressure cases with radiative cooling to 8000 K and 300 K, respectively, 150 Myr after the wind has hit the galaxy.  Note that the galaxy with cooling to 300 K has formed more high density clouds in the disk outskirts that are not stripped.  See \S \ref{sec-cool} for discussion.}
\label{fig:cool_image}
\end{figure*}

To better compare how an ICM wind would affect a galaxy with a lower cooling floor, we ran a simulation which is identical to PHRCW, but with radiative cooling to 300 K.  We find a few interesting results, shown in Figure \ref{fig:coolcomp}.  First, more gas is lost than in either PHRCW or PHNCW.  Gas is initially lost more slowly than in the other two cases, but does not slow down with fallback as in PHNCW.  Also, the final radius of the gas disk is similar to PHRCW, but it reaches this radius more slowly.  Finally, the amount of gas in each of the three density ranges is similar to that in PHRCW, but the two lower density regimes have a little less gas.  The highest density gas is the same in both runs throughout the simulations.  We also show (Figure \ref{fig:cool_image}) an image of each of the two cooling cases 150 Myr after the wind has hit the galaxy.

All of the differences in these two cases are explained by the fact that the galaxy with the lower radiative cooling floor fragments more than our standard cooled galaxy.  Because more of the gas is in dense knots, the initial onslaught of the wind does not strip gas as quickly.  Further, because gas that began in the outer regions of the disk survives, we see epicyclic variation in the radius, similar to our other simulations, resulting in an increase in the radius.  This is unlike PHRCW, in which all the gas that began in the outskirts is quickly stripped.  Because the clouds in the 300 K cooling case move to large radii, there is more time for them to be ablated as they drift back towards the center of the galaxy.  This explains why gas takes longer to be stripped, why more gas is stripped, and why the amount of mid-density gas decreases in this galaxy.  In the inner region of the disk, the larger amount of fragmentation means that more gas is stripped at early times from the central regions of the disk, because the gas disk is not smooth and cannot deflect the ICM wind.  In fact, we do not see gas fallback, as in the other high ram pressure cases, presumably because the wind pushes gas at all radii.  The longer period of gas stripping results both from ablation of the dense clouds in the center of the disk that are more exposed to the wind because of the larger amount of fragmentation, and because the dense clouds at large radii can be ablated more easily than their counterparts in the 8000 K case, which are closer to the galactic center.  

It is unclear what level of fragmentation is more realistic in observed galaxies.  Because of the lower Jeans length in the lower temperature floor runs, we resolve the resulting fragmentation less well and so prefer the runs with a higher minimum temperature.  In addition, it is not clear if the lower minimum temperature runs are physically more realistic as we do not include some effects such as small-scale turbulence, cosmic rays and magnetic fields, all of which may provide a source of effective pressure in low temperature regions.  While our simulations represent a sizable step forward in realism, it is clear that more work is required.

\section{Conclusions}

We have run a set of detailed galaxy simulations including radiative cooling to understand how a multiphase ISM interacts with an oncoming ICM wind.  We find:

1.  The total amount of gas mass loss is similar when comparing galaxies with and without radiative cooling.  However, the time over which the gas is stripped is shorter in the galaxies with radiative cooling.  The continued stripping of galaxies without radiative cooling could be Kelvin-Helmholtz stripping acting on the larger disk.

2.  The stripping radius is significantly smaller for the galaxy simulations with radiative cooling.  This is even true when accounting for the angular momentum loss in the ISM gas due to the interaction between the ISM and ICM gas (see section~\ref{sec-nowind}).

3. The morphology of the stripped disks are considerably different, with the cooling case producing a multiphase medium which results in many holes in the disk, while the case without cooling produces a coherent disk.  These holes allow ICM gas to stream through the galaxy disk, affecting the evolution of the dense clouds which remain.

4.  The amount of high density gas is affected differently depending on the amount of ram pressure.  When the ram pressure is strong, there is less gas at all densities, including the highest density.  This is because the low density gas is stripped, even from the inner regions of the galaxy, so there is less gas to feed molecular clouds.  With a weaker ram pressure, there is more of the highest density gas, although there is also less of the lower density gas.  A weaker wind strips less of the lower density gas, and increases the pressure on disk gas to create more high density gas.

This leads us to a picture in which radiative cooling results in both overdensities and underdensities at all radii in the galactic gas disk.  The underdensities are clearly stripped quickly throughout the disk, and the overdense regions that then have the wind streaming around them through the holes in the disk are stripped more rapidly than without a multiphase medium.  However, if a dense cloud is able to survive this initial onslaught, it loses angular momentum through its interaction with the non-rotating ICM wind and therefore settles toward a lower radius.  This is how the runs with cooling and a wind (the RCW cases) are stripped quickly and to a small radius while not losing much more gas than the galaxies without radiative cooling.  Also, the galaxies with radiative cooling seem to stop losing gas while those without cooling continue to lose gas through the entire simulation.  This may be because the Kelvin-Helmholtz instability is inefficient on small dense clouds but can affect the large cohesive disk in the no-cooling runs.  Allowing cooling to a lower temperature supports our picture that dense clouds cannot be directly stripped but can be ablated over time.  Although these simulations are arguably the most realistic stripping simulations performed to date, we have discussed in detail the limitations involved in our work in \S \ref{sec-discussion}.  

\acknowledgements

We acknowledge support from NSF grants AST-05-07161, AST-05-47823, and AST-06-06959, as well as computational resources from the National Center for Supercomputing Applications.  We thank Jacqueline van Gorkom, Jeff Kenney, Hugh Crowl, Airee Chung, Tomer Tal, and Anne Abramson for useful discussions, as well as Elizabeth Tasker for invaluable help setting up the initial conditions.  We would also like to thank our referee for editing and advice that greatly improved our paper.


\begin{appendix}

\section{The Nine Runs}

\subsection{PHNCW}

This galaxy begins with a gas radius of 21 kpc.  The surrounding ICM has a density and temperature listed in Table \ref{tbl-ICM}.  After about 67 Myr, an ICM wind with a Mach number of 3.5 and the density and temperature listed in Table \ref{tbl-ICM} hits the galaxy with a ram pressure of $6.4 \times 10^{-12}$ dynes cm$^{-3}$.  For about 50 Myr very little gas is stripped from the galaxy.  At that point, $2 \times 10^9$ M$_\odot$ is stripped from the disk in just over 50 Myr, although it remains gravitationally bound in the halo for another 150 Myr.  During this time, some of the gas that was stripped to more than 5 kpc above the disk falls back onto the galactic disk.  Once the gas that was initially stripped from the galaxy through ram pressure pushing has left the halo, gas that leaves the disk is unbound.  The total amount of gas mass loss in 900 Myr is $5.86 \times 10^9$ M$_\odot$.  Although there are strong fluctuations, the radius of remaining gas settles to about 11 kpc.  Although all the gas in this galaxy is low density, by making finer density cuts we find that the gas with the highest densities, in this case between $10^{-24}$ and $10^{-23}$ g cm$^{-3}$, stays nearly constant, with 80\% of the stripped gas coming from lower density gas.

\subsection{PHRCNW}\label{PHRCNW}

This galaxy begins with a gas radius of 25 kpc (as do all of the galaxies in simulations that include radiative cooling).  Throughout this run the surrounding ICM has the density and temperature of the post-shock ICM when there is a wind in the high ram pressure cases.  We chose the hotter ICM because of it's longer cooling time.  Initially the radius seems to fluctuate with the epicyclic period ($\sim 500$ Myr), however, cooling creates both over densities and underdensities in the disk.  This allows some of the non-rotating ICM to fall into holes in the disk and to slow the disk's rotation, leading to a drop in the disk radius through the loss of angular momentum.  This gas infall is also seen by considering the total amount of gas in the disk region:  the galaxy gains $3.8 \times 10^8$ M$_\odot$ over the 950 Myr run.  High density gas forms about 100 Myr into the galaxy's evolution, with the highest density gas, above $10^{-20}$ g cm$^{-3}$, forming closer to 150 Myr.  Throughout the run the galaxy gains gas with densities above $10^{-22}$ g cm$^{-3}$, loses less dense gas as it turns into more dense gas.  Although we would need to extend this run to be sure the radius would stay constant, the radius settles between 11 and 12 kpc in the final 200 Myr of the run.

\subsection{PHRCW}

As mentioned above, because this galaxy includes radiative cooling, it begins with a radius of 25 kpc.  The surrounding density and temperature are the same as run PHNCW, shown in Table \ref{tbl-ICM}.  After about 190 Myr, the same ICM wind hits this galaxy as hits the galaxy in PHNCW.  Very little gas is lost for a few tens of Myrs, at which point $5 \times 10^9$ M$_\odot$ is lost from the galaxy in under 150 Myr.  $3 \times 10^9$ M$_\odot$ is immediately gravitationally unbound from the galaxy, and almost all of the remaining $2 \times 10^9$ M$_\odot$ leaves the halo within 400 Myr.  A small amount of this gas falls back onto the galactic plane.  After the gas immediately removed from the galactic disk by ram pressure pushing is no longer gravitationally bound, about 550 Myr after the wind first hits the galaxy, gas loss precedes at about 1 M$_\odot$ yr$^{-1}$ for the final 250 Myr we continue the run.  The total gas mass loss is about $5.8 \times 10^9$ M$_\odot$.  As in the PHRCNW run, the radius of the disk falls before the ICM wind hits the galaxy.  Once the wind hits the galaxy the gas is stripped to a radius below 9 kpc within 250 Myr.  The radius after 800 Myr of ram pressure stripping is at or below 7 kpc.  As in PHRCNW, the galaxy loses lower density gas (below $10^{-22}$ g cm$^{-3}$) throughout the run.  Although the galaxy does increase all it's higher density gas, gas with densities between $10^{-20}$ and $10^{-22}$ g cm$^{-3}$ remains constant after the ICM wind has been hitting the galaxy for 350 Myr.  Also, the highest density gas does not appear until 350 Myr into the simulation, about 150 Myr after the ICM wind hits the galaxy.

\subsection{PMNCW}

This galaxy begins with a radius of 21 kpc.  The surrounding ICM density and temperature is listed in Table \ref{tbl-ICM}.  After the galaxy has evolved for about 80 Myr, an ICM wind hits the galaxy with a ram pressure of $1 \times 10^{-12}$ dynes cm$^{-2}$, and density and temperature listed in Table \ref{tbl-ICM}.  The total amount of gas loss in the more than 900 Myr in which the ICM wind hits the galaxy is less than $9.5 \times 10^8$ M$_\odot$.  The mass loss rate for the first 100 Myr is less than 0.5 M$_\odot$ yr$^{-1}$, and increases until in the final 100 Myr it is a little over 2 M$_\odot$ yr$^{-1}$.  The radius fluctuates, but seems to settle at about 19 kpc.  In this case, all the gas that is stripped is less dense than $10^{-24}$ g cm$^{-3}$.

\subsection{PMRCNW}

This galaxy begins with a radius of 21 kpc.  This surrounding ICM density and temperature are the middle post-shock values listed in Table \ref{tbl-ICM}.  Again the radius seems to vary due to a combination of epicyclic variation and mixing from non-rotating ICM gas.   Less ICM gas is accreted in this case than in PHRCNW, $2.1 \times 10^8$ M$_\odot$, which can be expected since the ICM pressure is lower.  As in the higher ICM pressure case, PHRCNW, throughout the run the galaxy gains gas with densities above $10^{-22}$ g cm$^{-3}$, and loses gas less dense than $10^{-22}$ g cm$^{-3}$.  However, there is less gas with densities greater than $10^{-20}$ g cm$^{-3}$ in this run than in PHRCNW, and it does not form until about 250 Myr.  The radius remains at about 15.5 kpc for the final 400 Myr of this run.

\subsection{PMRCW}

This galaxy begins with a radius of 25 kpc.  The surrounding ICM density and temperature values are the same as in PMNCW.  After the galaxy has evolved for about 275 Myr, the ICM wind hits the galaxy with a ram pressure of $1 \times 10^{-12}$ dynes cm$^{-2}$.  For 100 Myr little gas is lost, and then in less than 350 Myr after the wind first hits the galaxy the majority of the gas is stripped, $1.23 \times 10^9$ M$_\odot$ of a total loss of $1.3 \times 10^9$ M$_\odot$.  The radius falls to 21 kpc before the ICM wind hits the galaxy, and then begins to fall again about 230 Myr after the wind first hits the galaxy.  The final radius, just below 11 kpc, is reached in slightly less than 900 Myr after the wind hits the galaxy and is constant for the final 200 Myr of the simulation.  The galaxy loses lower density gas (below $10^{-22}$ g cm$^{-3}$) throughout the run.  Although the galaxy does increase it's higher density gas, gas with densities between $10^{-20}$ and $10^{-22}$ g cm$^{-3}$ remains constant after the ICM wind has been hitting the galaxy for 700 Myr.  The highest density gas, above $10^{-20}$ g cm$^{-3}$, is first formed 450 Myr into the simulation, or 175 Myr after the higher pressure ICM hits the galaxy.  

\subsection{PLNCW}

This galaxy begins with a radius of 21 kpc.  The surrounding ICM density and temperature are listed in the bottom pre-shock rows of Table \ref{tbl-ICM}.  After the galaxy has evolved for about 77 Myr, the ICM wind hits the galaxy with the post-shock values in Table \ref{tbl-ICM} and a ram pressure of $6.4 \times 10^{-13}$ dynes cm$^{-2}$.  The total gas loss is $5.8 \times 10^8$ M$_\odot$.  The gas loss is spread throughout the entire time the wind hits the galaxy, with the fastest gas loss rate less than 2 M$_\odot$ yr$^{-1}$.  The gas loss rate does not constantly ascend or descend.  The radius of the disk oscillates with a period of about 370 Myr, the epicyclic period of a disk with a radius of 18 kpc, the Gunn \& Gott stripping radius for this case.  The oscillations are too large, however, to make any claims about a stripping radius.  Once again, in this case, all the gas that is stripped is less dense than $10^{-24}$ g cm$^{-3}$.

\subsection{PLRCNW}

This galaxy begins with a radius of 25 kpc.  The surrounding ICM density and temperature are listed in the bottom post-shock rows of Table \ref{tbl-ICM}.  In this case radial variation seems to be mostly due to epicycles, with mixing perhaps having the effect of shifting the oscillations to lower radii.  This galaxy gains less than $5 \times 10^7$ M$_\odot$ in the entire 1.2 Myr of the run.  As in both of the other RCNW cases, throughout the run the galaxy gains gas with densities above $10^{-22}$ g cm$^{-3}$, and loses gas less dense than $10^{-23}$ g cm$^{-3}$.  However, there is less gas with densities greater than $10^{-20}$ g cm$^{-3}$ in this run than in the runs with higher ICM pressure, although it does form at about 300 Myr, nearly the same time as in PMRCNW.    

\subsection{PLRCW}

This galaxy begins with a radius of 25 kpc.  The surrounding ICM density and temperature are the same pre-shock values as in PLNCW.  After the galaxy has evolved for 271 Myr, the ICM wind hits the galaxy with a ram pressure of $6.4 \times 10^{-13}$ dynes cm$^{-2}$.  As in PMRCW, there is little change in gas mass for the first 100 Myr after the wind hits the galaxy, and then the galaxy loses $7.34 \times 10^8$ M$_\odot$, of total gas mass loss of $8.5 \times 10^8$ M$_\odot$, in less than 350 Myr.  Even with this small gas mass loss rate from the disk of less than 2 M$_\odot$ yr$^{-1}$, some small (less than $2 \times 10^8$ M$_\odot$) amount of the stripped gas remains in the halo until 600 Myr after the wind has hit the galaxy.  After the initial gas loss, the removal rate is small and continues to approach zero.  Again the radius of the galaxy decreases in a smooth curve and is hit by the wind before it would rise in an epicyclic pattern.  Once the galaxy is stripped, the radius decreases to between 15 and 14 kpc, and seems to hold steady there for the last 600 Myr of the simulation.  The galaxy loses lower density gas (below $10^{-22}$ g cm$^{-3}$) throughout the run.  The galaxy increases it's higher density gas (denser than $10^{-22}$ g cm$^{-3}$) throughout the run.  The highest density gas, above $10^{-20}$ g cm$^{-3}$, is first formed 500 Myr into the simulation, or 225 Myr after the higher pressure ICM hits the galaxy.

\end{appendix}


\begin{thebibliography}

\bibitem[Barnes \& Hernquist(1991)]{1991ApJ...370L..65B} Barnes, J.~E., \& Hernquist, L.~E.\ 1991, \apjl, 370, L65 

\bibitem[Bekki(1999)]{1999ApJ...510L..15B} Bekki, K.\ 1999, \apjl, 510, L15 

\bibitem[Bekki(1998)]{1998ApJ...502L.133B} Bekki, K.\ 1998, \apjl, 502, L133 

\bibitem[Bigiel et al.(2008)]{2008AJ....136.2846B} Bigiel, F., Leroy, A., Walter, F., Brinks, E., de Blok, W.~J.~G., Madore, B., z\& Thornley, M.~D.\ 2008, \aj, 136, 2846

\bibitem[Blitz 
\& Shu(1980)]{1980ApJ...238..148B} Blitz, L., \& Shu, F.~H.\ 1980, \apj, 238, 148 

\bibitem[Bryan(1999)]{Bryan1999}Bryan,G.L. Comp. Phys. and Eng. 1999, 1:2, p.

\bibitem[Burkert(1995)]{1995ApJ...447L..25B} Burkert, A.\ 1995, \apjl, 447, L25 

\bibitem[Butcher \& Oemler(1978)]{1978ApJ...219...18B} Butcher, H., \& Oemler, A., Jr.\ 1978, \apj, 219, 18 

\bibitem[Byrd \& Valtonen(1990)]{1990ApJ...350...89B} Byrd, G., \& Valtonen, M.\ 1990, \apj, 350, 89 

\bibitem[Clemens et al.(2001)]{2001Ap&SS.276..451C} Clemens, M., Alexander, P., \& Green, D.\ 2001, \apss, 276, 451 

\bibitem[Clemens et al.(2000)]{2000MNRAS.312..236C} Clemens, M.~S., 
Alexander, P., \& Green, D.~A.\ 2000, \mnras, 312, 236 

\bibitem[Crowl et al.(2005)]{2005AJ....130...65C} Crowl, H.~H., Kenney, 
J.~D.~P., van Gorkom, J.~H., \& Vollmer, B.\ 2005, \aj, 130, 65 

\bibitem[Crutcher(1991)]{1991IAUS..147...61C} Crutcher, R.~M.\ 1991, 
Fragmentation of Molecular Clouds and Star Formation, 147, 61 

\bibitem[Dressler(1980)]{1980ApJ...236..351D} Dressler, A.\ 1980, \apj, 236, 351 

\bibitem[Fujita \& Nagashima(1999)]{1999ApJ...516..619F} Fujita, Y., \& Nagashima, M.\ 1999, \apj, 516, 619 

\bibitem[Giovanelli \& Haynes(1985)]{1985ApJ...292..404G} Giovanelli, R., \& Haynes, M.~P.\ 1985, \apj, 292, 404 

\bibitem[Gunn \& Gott(1972)]{1972ApJ...176....1G} Gunn, J.~E., \& Gott, J.~R.~I.\ 1972, \apj, 176, 1 

\bibitem[Joung \& Mac Low(2006)]{2006ApJ...653.1266J} Joung, M.~K.~R., \& Mac Low, M.-M.\ 2006, \apj, 653, 1266 

\bibitem[Hartmann(2003)]{2003ApJ...585..398H} Hartmann, L.\ 2003, \apj, 585, 398 

\bibitem[Hashimoto et al.(1998)]{1998ApJ...499..589H} Hashimoto, Y., 
Oemler, A.~J., Lin, H., \& Tucker, D.~L.\ 1998, \apj, 499, 589 

\bibitem[Hernquist(1993)]{1993ApJS...86..389H} Hernquist, L.\ 1993, \apjs, 86, 389 

\bibitem[Hubble \& Humason(1931)]{1931ApJ....74...43H} Hubble, E., \& Humason, M.~L.\ 1931, \apj, 74, 43 

\bibitem[Kenney \& Young(1986)]{1986ApJ...301L..13K} Kenney, J.~D., \& Young, J.~S.\ 1986, \apjl, 301, L13 

\bibitem[Kenney 
\& Young(1989)]{1989ApJ...344..171K} Kenney, J.~D.~P., \& Young, J.~S.\ 1989, \apj, 344, 171 

\bibitem[Kenney \& Koopmann(1999)]{1999AJ....117..181K} Kenney, J.~D.~P., \& Koopmann, R.~A.\ 1999, \aj, 117, 181 

\bibitem[Kenney et al.(2004)]{2004AJ....127.3361K} Kenney, J.~D.~P., van 
Gorkom, J.~H., \& Vollmer, B.\ 2004, \aj, 127, 3361 

\bibitem[Kronberger et al.(2008)]{2008A&A...481..337K} Kronberger, T., Kapferer, W., Ferrari, C., Unterguggenberger, S., \& Schindler, S.\ 2008, \aap, 481, 337 


\bibitem[Larson(2003)]{2003RPPh...66.1651L} Larson, R.~B.\ 2003, Reports of 
Progress in Physics, 66, 1651 

\bibitem[Leroy et al.(2008)]{2008arXiv0810.2556L} Leroy, A.~K., Walter, F., 
Brinks, E., Bigiel, F., de Blok, W.~J.~G., Madore, B., 
\& Thornley, M.~D.\ 2008, arXiv:0810.2556 

\bibitem[Men et al.(2008)]{2008A&A...486..819M} Men, H., Ferri{\`e}re, K., \& Han, J.~L.\ 2008, \aap, 486, 819 

\bibitem[Miyamoto \& Nagai(1975)]{1975PASJ...27..533M} Miyamoto, M., \& Nagai, R.\ 1975, \pasj, 27, 533 

\bibitem[Moore et al.(1996)]{1996Natur.379..613M} Moore, B., Katz, N., Lake, G., Dressler, A., \& Oemler, A.\ 1996, \nat, 379, 613 

\bibitem[Mori \& Burkert(2000)]{2000ApJ...538..559M} Mori, M., \& Burkert, A.\ 2000, \apj, 538, 559 

\bibitem[Nakanishi et al.(2006)]{2006ApJ...651..804N} Nakanishi, H., et 
al.\ 2006, \apj, 651, 804 

\bibitem[Norman \& Bryan(1999)]{Norman1999} Norman, M.~L.~\&
Bryan, G.~L.\ 1999, ASSL Vol.~240: Numerical Astrophysics, 19

\bibitem[Nulsen(1982)]{1982MNRAS.198.1007N} Nulsen, P.~E.~J.\ 1982, \mnras, 
198, 1007 

\bibitem[O'Shea et al.(2004)]{OShea2004} O'Shea, B.~W., Bryan, G., Bordner, J., Norman, M.~L., Abel, T., Harkness, R., \& Kritsuk, A.\ 2004, ArXiv Astrophysics e-prints, astro-ph/0403044 

\bibitem[Oemler(1974)]{1974ApJ...194....1O} Oemler, A.~J.\ 1974, \apj, 194, 1 

\bibitem[Poggianti et al.(2004)]{2004ApJ...601..197P} Poggianti, B.~M., 
Bridges, T.~J., Komiyama, Y., Yagi, M., Carter, D., Mobasher, B., Okamura, 
S., \& Kashikawa, N.\ 2004, \apj, 601, 197 

\bibitem[Quilis et al.(2000)]{2000Sci...288.1617Q} Quilis, V., Moore, B., 
\& Bower, R.\ 2000, Science, 288, 1617 

\bibitem[Roediger \& Hensler(2005)]{2005A&A...433..875R} Roediger, E., \& Hensler, G.\ 2005, \aap, 433, 875 

\bibitem[Roediger \& Br{\"u}ggen(2007)]{2007MNRAS.380.1399R} Roediger, E., \& Br{\"u}ggen, M.\ 2007, \mnras, 380, 1399 

\bibitem[Roediger \& Br{\"u}ggen(2006)]{2006MNRAS.369..567R} Roediger, E., \& Br{\"u}ggen, M.\ 2006, \mnras, 369, 567 

\bibitem[Sarazin \& White(1987)]{cooling} Sarazin, C. \& White, 1987, Apj, 320, 32

\bibitem[Schulz \& Struck(2001)]{2001MNRAS.328..185S} Schulz, S., \& Struck, C.\ 2001, \mnras, 328, 185 

\bibitem[Tasker 
\& Bryan(2006)]{2006ApJ...641..878T} Tasker, E.~J., \& Bryan, G.~L.\ 2006, \apj, 641, 878 

\bibitem[Tasker 
\& Bryan(2008)]{2008ApJ...673..810T} Tasker, E.~J., \& Bryan, G.~L.\ 2008, \apj, 673, 810 

\bibitem[Tonnesen et al.(2007)]{2007ApJ...671.1434T} Tonnesen, S., Bryan, 
G.~L., \& van Gorkom, J.~H.\ 2007, \apj, 671, 1434 

\bibitem[Trachternach et al.(2008)]{2008AJ....136.2720T} Trachternach, C., 
de Blok, W.~J.~G., Walter, F., Brinks, E., 
\& Kennicutt, R.~C.\ 2008, \aj, 136, 2720 

\bibitem[Tran et al.(2005)]{2005ApJ...619..134T} Tran, K.-V.~H., van Dokkum, P., Illingworth, G.~D., Kelson, D., Gonzalez, A., \& Franx, M.\ 2005, \apj, 619, 134 

\bibitem[van Gorkom(2004)]{2004cgpc.symp..305V} van Gorkom, J.~H.\ 2004, 
Clusters of Galaxies: Probes of Cosmological Structure and Galaxy 
Evolution, 305 

\bibitem[Vollmer et al.(2001)]{2001A&A...374..824V} Vollmer, B., Braine, J., Balkowski, C., Cayatte, V., \& Duschl, W.~J.\ 2001, \aap, 374, 824

\bibitem[Vollmer et al.(2005)]{2005A&A...439..921V} Vollmer, B., 
Huchtmeier, W., \& van Driel, W.\ 2005, \aap, 439, 921 

\bibitem[Vollmer et al.(2005)]{2005A&A...441..473V} Vollmer, B., Braine, J., Combes, F., \& Sofue, Y.\ 2005, \aap, 441, 473 

\end{thebibliography}
\end{document}